\documentclass[useAMS,usenatbib,fleqn,a4paper]{mn2e}
\usepackage{graphicx}
\usepackage{setspace}
\usepackage{psfig}
\usepackage{natbib}
\usepackage{epsfig}
\usepackage{deluxetable} 
\usepackage{url}
\usepackage{rotating}
\usepackage{amsmath}
\usepackage{float}
\usepackage{placeins}

\def\aap{A\&A}%
\def\mnras{MNRAS}%
\def\na{New A}%
\def\nat{Nature}%
\title[MWL behaviour of CTA~102 during 2013--2017]{Investigating the multiwavelength behaviour of the flat spectrum radio quasar CTA~102 during 2013--2017}
\author[D'Ammando, Raiteri, Villata, et al.]{F.~D'Ammando$^{1}$\thanks{E-mail: dammando@ira.inaf.it}, 
C.~M.~Raiteri$^{2}$, 
M.~Villata$^{2}$,
J.~A.~Acosta-Pulido$^{3,4}$,
I.~Agudo$^{5}$,
\newauthor A.~A.~Arkharov$^{6}$,
R.~Bachev$^{7}$,
G.~V.~Baida$^{8}$,
E.~Ben\'itez$^{9}$,
G.~A.~Borman$^{8}$,
\newauthor W.~Boschin$^{10,3,4}$,
V.~Bozhilov$^{11}$,
M.~S.~Butuzova$^{8}$,
P.~Calcidese$^{12}$,
M.~I.~Carnerero$^{2}$,
\newauthor D.~Carosati$^{13,10}$,
C.~Casadio$^{14,5}$,
N.~Castro-Segura$^{4,15}$,
W.-P.~Chen$^{16}$,
G.~Damljanovic$^{17}$, 
\newauthor A.~Di~Paola$^{18}$,
J.~Echevarr\'ia$^{9}$,
N.~V.~Efimova$^{6}$, 
Sh.~A.~Ehgamberdiev$^{19}$,
C.~Espinosa$^{9}$,
\newauthor A.~Fuentes$^{5}$,
A.~Giunta$^{18}$,
J.~L.~G\'omez$^{5}$,
T.~S.~Grishina$^{20}$,
M.~A.~Gurwell$^{21}$,
D.~Hiriart$^{9}$,
\newauthor H.~Jermak$^{22}$,
B.~Jordan$^{23}$,
S.~G.~Jorstad$^{24,20}$,
M.~Joshi$^{24}$,
G.~N.~Kimeridze$^{27}$,
\newauthor E.~N.~Kopatskaya$^{20}$,
K.~Kuratov$^{25,26}$,
O.~M.~Kurtanidze$^{27,28,29,30}$,
S.~O.~Kurtanidze$^{27}$,
\newauthor A.~L\"ahteenm\"aki$^{31,32}$, 
V.~M.~Larionov$^{20,6}$,
E.~G.~Larionova$^{20}$,
L.~V.~Larionova$^{20}$,
\newauthor C.~L\'azaro$^{3,4}$,
C.~S.~Lin$^{16}$,
M.~P.~Malmrose$^{24}$,
A.~P.~Marscher$^{24}$,
K.~Matsumoto$^{33}$,
\newauthor B.~McBreen$^{34}$,
R.~Michel$^{9}$,
B.~Mihov$^{7}$,
M.~Minev$^{11}$,
D.~O.~Mirzaqulov$^{19}$,
S.~N.~Molina$^{5}$,
\newauthor J.~W.~Moody$^{35}$,
D.~A.~Morozova$^{20}$,
S.~V.~Nazarov$^{8}$,
A.~A.~Nikiforova$^{20,6}$,
\newauthor M.~G.~Nikolashvili$^{27}$,
J.~M.~Ohlert$^{36,37}$,
N.~Okhmat$^{8}$,
E.~Ovcharov$^{9}$,
F.~Pinna$^{3,4}$,
\newauthor T.~A.~Polakis$^{38}$,
C.~Protasio$^{3,4}$,
T.~Pursimo$^{39}$,
F.~J.~Redondo-Lorenzo$^{3,4}$,
N.~Rizzi$^{40}$,
\newauthor G.~Rodriguez-Coira$^{41}$,
K.~Sadakane$^{33}$,
A.~C.~Sadun$^{42}$,
M.~R.~Samal$^{16}$,
S.~S.~Savchenko$^{20}$,
\newauthor E.~Semkov$^{7}$,
L.~Sigua$^{27}$,
B.~A.~Skiff$^{43}$,
L.~Slavcheva-Mihova$^{7}$,
P.~S.~Smith$^{44}$,
I.~A.~Steele$^{22}$,
\newauthor A.~Strigachev$^{7}$,
J.~Tammi$^{31}$,
C.~Thum$^{45}$,
M.~Tornikoski$^{31}$,
Yu.~V.~Troitskaya$^{20}$,
\newauthor I.~S.~Troitsky$^{20}$,
A.~A.~Vasilyev$^{20}$,
O.~Vince$^{17}$ (the WEBT Collaboration), T.~Hovatta$^{46,47}$,
\newauthor S.~Kiehlmann$^{48}$,
W.~Max-Moerbeck$^{49}$,
A.~C.~S.~Readhead$^{48}$,
R.~Reeves$^{50}$, T.~J.~Pearson$^{48}$ \newauthor (the OVRO Team), T.~Mufakharov$^{51,52}$,
Yu.~V.~Sotnikova$^{53}$, and M.~G.~Mingaliev$^{52,53}$
}

\date{Accepted 2019 October 02}
\pubyear{2019}

\begin{document}

\label{firstpage}
\pagerange{\pageref{firstpage}--\pageref{lastpage}}

\maketitle

\begin{abstract}
\small
We present a multiwavelength study of the flat-spectrum radio quasar CTA~102 during 2013--2017. We use radio-to-optical data obtained by the Whole Earth Blazar Telescope, 15 GHz data from the Owens Valley Radio Observatory, 91 and 103 GHz data from the Atacama Large Millimeter Array, near-infrared data from the Rapid Eye Monitor telescope, as well as data from the {\em Swift} (optical-UV and X-rays) and {\em Fermi} ($\gamma$ rays) satellites to study flux and spectral variability and the correlation between flux changes at different wavelengths. Unprecedented $\gamma$-ray flaring activity was observed during 2016 November--2017 February, with four major outbursts. A peak flux of (2158 $\pm$ 63)$\times$10$^{-8}$ ph cm$^{-2}$ s$^{-1}$, corresponding to a luminosity of (2.2 $\pm$ 0.1)$\times$10$^{50}$ erg s$^{-1}$, was reached on 2016 December 28. These four $\gamma$-ray outbursts have corresponding events in the near-infrared, optical, and UV bands, with the peaks observed at the same time. A general agreement between X-ray and $\gamma$-ray activity is found. The $\gamma$-ray flux variations show a general, strong correlation with the optical ones with no time lag between the two bands and a comparable variability amplitude. This $\gamma$-ray/optical relationship is in agreement with the geometrical model that has successfully explained the low-energy flux and spectral behaviour, suggesting that the long-term flux variations are mainly due to changes in the Doppler factor produced by variations of the viewing angle of the emitting regions. The difference in behaviour between radio and higher energy emission would be ascribed to different viewing angles of the jet regions producing their emission.  
 
\end{abstract} 


\section{Introduction}

Blazars are an extreme class of active galactic nuclei (AGN) whose bright and
violently variable non-thermal radiation across the entire electromagnetic
spectrum is ascribed to the presence of a collimated relativistic jet closely aligned to our line of sight
\citep[e.g.,][]{blandford78}. This peculiar setting implies a strong amplification of the rest-frame radiation because of Doppler boosting, together with a contraction of the variability time-scales, and a blueshift of the frequencies. 

The relativistic jets of blazars are able to transport a huge amount of power away from the central engine in the form of radiation, kinetic energy, and magnetic fields. When this power is dissipated, the particles emit the observed radiation, showing the typical double-hump spectral energy distribution (SED) of blazars. The first peak of the SED, usually observed between radio and X-rays, is due to the synchrotron radiation from relativistic electrons, while the second peak, usually observed from X-ray up to TeV energies, is commonly interpreted as inverse Compton (IC) scattering of seed photons, either internal or external to the jet, by highly relativistic electrons. However, the nature of this second hump is a controversial issue and other models involving hadronic and lepto-hadronic processes have been proposed \citep[e.g.,][]{boettcher13}.

Blazars are traditionally divided into flat-spectrum radio quasars (FSRQ) and BL Lac objects (BL Lacs), based on the presence or not, respectively, of broad
emission lines (i.e., equivalent width $>$ 5 \AA) in their optical and UV spectrum \citep[e.g.,][]{stickel91}. Recently, a new classification was proposed based on the luminosity of the broad-line region (BLR) in Eddington luminosity \citep{ghisellini11}: sources with $L_{BLR}$/$L_{Edd}$ higher or lower than 5$\times$10$^{-4}$ being classified as FSRQ or BL Lacs, respectively, in agreement with a transition of the accretion regime from efficient to inefficient between the two classes.  

Blazar emission shows strong and unpredictable variability over all the
electromagnetic spectrum, from the radio band to $\gamma$ rays, with time-scales
ranging from minutes to years. Long-term observations of blazars during
different activity states provide an ideal laboratory for investigating the
emission mechanisms at work in this class of sources.
In this paper we present multifrequency observations of the blazar CTA~102
during 2013--2017. CTA~102 (also known as 4C $+$11.69) is an FSRQ at redshift $z = 1.037$ \citep{schmidt65}.  
Flaring activity in the optical band has been observed from this source in 1978
\citep{pica88}, 1996 \citep{katajainen00}, and 2004 \citep{osterman09}. However, simultaneous $\gamma$-ray observations were not
available for those events. The source was detected for the first time in
$\gamma$ rays by the {\em Compton Gamma Ray Observatory} in 1992 with both
the EGRET \citep{hartman99} and COMPTEL \citep{blom95} instruments. Unfortunately,
no optical observations were available during the $\gamma$-ray detection \citep{villata97}.
On the other hand, during the {\em Fermi} era a remarkable outburst was
simultaneously observed
in 2012 September--October in near-infrared (near-IR) and optical bands by the Whole Earth Blazar Telescope\footnote{http://www.oato.inaf.it/blazars/webt} (WEBT) and $\gamma$ rays by the Large Area Telescope (LAT) on board {\em Fermi Gamma-ray Space Telescope}. Correlated variability in the two energy bands suggested a co-spatial
origin of the optical and $\gamma$-ray emitting regions during the flaring activity \citep{lar16}. 

\noindent In 2016 November, CTA~102 entered a new very-high-activity state in $\gamma$ rays, as observed by {\em Fermi}-LAT, reaching a daily flux higher than 1$\times$10$^{-5}$ ph cm$^{-2}$ s$^{-1}$ on 2016 December 16
\citep{ciprini16}. This flaring activity continued for a few weeks in $\gamma$
rays \citep[e.g.,][]{bulgarelli16, xu16}. A significant increase of activity was observed over the entire electromagnetic spectrum
\citep[e.g.,][]{calcidese16,ojha16,righini16}. In particular, an 
extreme optical and near-IR outburst occurred in 2016 December, with a
brightness increase up to six magnitudes with respect to the faint state of
the source \citep{rai17}. In \citet{rai17} we explained the flux and spectral variations in optical, near-IR, and radio bands by means of an inhomogeneous curved jet with different jet regions changing their orientation, and hence their Doppler factors, in time. Alternative theoretical scenarios have been proposed to explain the 2016--2017 flaring behaviour of CTA~102. According to \citet{casadio19} the outburst was produced by a superluminal component crossing a recollimation shock, while for \citet{zacharias17, zacharias19} it was due to ablation of a gas cloud penetrating the relativistic jet in a leptonic or hadronic scenario. 

The radio-to-optical and $\gamma$-ray emission are produced by two different mechanisms (i.e.~synchrotron and IC emission in leptonic models), although related to the same relativistic electron population. Therefore, the $\gamma$-ray variability can be used as a further test to verify the geometrical model that we proposed to explain the low-energy flux variability in CTA 102 during 2013--2017. In the geometrical scenario, the $\gamma$-ray and optical radiation are produced in the same jet region, therefore the $\gamma$-ray and optical fluxes undergo the same Doppler beaming and should be linearly correlated. 

In this paper we present a multiwavelength analysis of the CTA~102 emission from radio to $\gamma$ rays between 2013 January 1 and 2017 February 9, in particular during the bright flaring activity occurred during 2016 November--2017 February. The radio-to-optical observations performed in the framework of a campaign led by the WEBT, already presented in \citet{rai17}, are complemented by the Atacama Large millimeter/Submillimeter Array (ALMA) at 91 and 103 GHz, the Owens Valley Radio Observatory (OVRO) data at 15 GHz, the Rapid Eye Mount (REM) near-IR data, and a detailed analysis of data collected by the {\em Neil Gehrels Swift Observatory} (optical--UV and X rays) and {\em Fermi} ($\gamma$ rays) satellites. The data set used in this paper is the richest in terms of number of data points and broad-band coverage presented in literature for the period considered here.

Sun constraints prevented us to have observations from optical and near-infrared WEBT observatories and {\em Swift} satellite after 2017 February 9, not allowing us to investigate the connection between the $\gamma$-ray flaring activity observed in 2017 March--April \citep[see e.g.,][]{shukla18} and the emission from near-IR to X-rays. After that period the infrared-to-X-ray coverage is insufficient to adequately test the geometrical model and to investigate the connection between low-energy and $\gamma$-ray emission.

The paper is organized as follows. In Section~\ref{FermiData} and \ref{SwiftData} we present {\it  Fermi}-LAT and {\em Swift} data analysis and results, respectively, whereas in Section~\ref{RadioOptical} we report on the radio-to-optical observations. Multifrequency flux and spectral variability are discussed in Sections~\ref{Multifrequency} and \ref{SpectralVariability}, respectively. The application of the geometrical model by \citet{rai17} to the $\gamma$-ray, optical, and radio variability is discussed in Section~\ref{geometry}. We discuss the previous results and draw our conclusions in Section~\ref{Summary}. Throughout this paper, we assume the following cosmology: $H_{0} = 71\; {\rm km \; s^{-1} \; Mpc^{-1}}$, $\Omega_{\rm M} = 0.27$, and $\Omega_{\rm \Lambda} = 0.73$ in a flat Universe \citep{planck16}. \\

\section{{\em Fermi}-LAT Data: analysis and results}\label{FermiData}

\begin{figure}
\centering
\includegraphics[width=8.5cm, height=8.5cm, keepaspectratio]{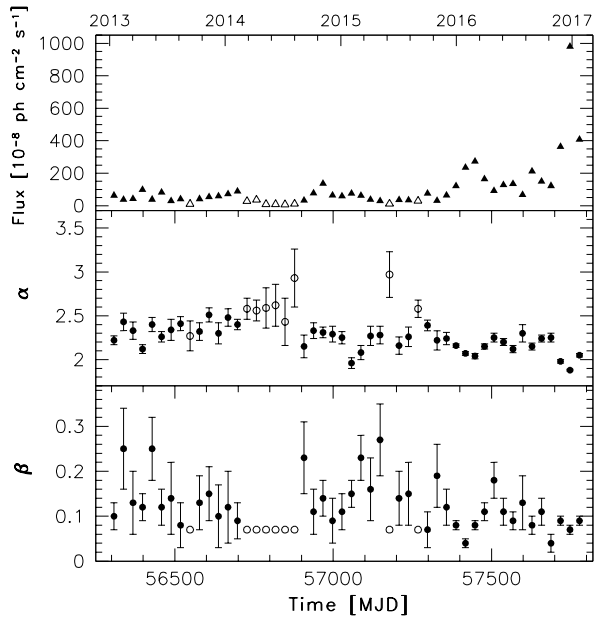}
\caption{Integrated flux light curve of CTA~102 (upper panel), spectral slope (middle panel), curvature parameter (bottom panel) obtained in the 0.1--300 GeV energy range during 2013 January 1--2017 February 9 (MJD 56293--57793) with 30-d time bins. The open symbols refer to results obtained with $\beta$ fixed to 0.07 (see the text for details).}
\label{Fig1}
\end{figure}

The {\em Fermi}-LAT  is a pair-conversion telescope operating from 20 MeV to
$>$ 300 GeV. Further details about the {\em Fermi}-LAT are given in
\citet{atwood09}.
 
\noindent The LAT data used in this paper were collected from 2013
January 1 (MJD 56293) to 2017 February 9 (MJD 57793). During this time, the
LAT instrument operated almost entirely in survey mode. The Pass 8 data
\citep{atwood13}, based on a complete and improved revision of the entire LAT
event-level analysis, were used. The analysis was performed with the
\texttt{ScienceTools} software package version v11r5p3.
Only events belonging to the `Source' class (\texttt{evclass=128}, \texttt{evtype=3}) were used. We selected only events within a maximum zenith angle of 90 deg to reduce contamination from the Earth limb $\gamma$ rays, which are produced by cosmic rays interacting with the upper atmosphere. 
The spectral analysis was performed with the instrument response functions \texttt{P8R2\_SOURCE\_V6} using a binned maximum-likelihood method implemented in the Science tool \texttt{gtlike}. Isotropic (`iso\_source\_v06.txt') and Galactic diffuse emission (`gll\_iem\_v06.fit') components were used to model the background \citep{acero16}\footnote{http://fermi.gsfc.nasa.gov/ssc/data/access/lat/\\BackgroundModels.html}. The normalization of both components was allowed to vary freely during the spectral fitting.

We analysed a region of interest of $20^{\circ}$ radius centred at the location of CTA~102. We evaluated the significance of the $\gamma$-ray signal from the source by means of a maximum-likelihood test statistic (TS) defined as TS = 2$\times$(log$L_1$ - log$L_0$), where
$L$ is the likelihood of the data given the model with ($L_1$) or without ($L_0$) a point source at the position of CTA~102 \citep[e.g.,][]{mattox96}. The source model used in \texttt{gtlike} includes all the point sources from the 3FGL catalogue that fall within $30^{\circ}$ of CTA~102. We also included new candidates within $10^{\circ}$ of CTA~102 from a preliminary eight-year point source list (FL8Y\footnote{https://fermi.gsfc.nasa.gov/ssc/data/access/lat/fl8y/}). The spectra of these sources were parametrized by a power law (PL), a log-parabola (LP), or a super exponential cut-off, as in the catalogues. 

A first maximum-likelihood analysis was performed over the whole period to remove from the model the sources having TS $< 25$. A second maximum-likelihood analysis was performed on the updated source model. In the fitting
procedure, the normalization factors and the spectral parameters of the
sources lying within 10$^{\circ}$ of CTA~102 were left as free parameters. For
the sources located between 10$^{\circ}$ and 30$^{\circ}$ from our target, we
kept the normalization and the spectral shape parameters fixed to the values
from the 3FGL catalogue.
 
Integrating over 2013 January 1--2017 February 9 the fit with an LP model, $dN/dE \propto$ $E/E_{0}^{-\alpha-\beta \, \log(E/E_0)}$, as in the 3FGL
and FL8Y catalogues, results in TS = 125005 in the 0.1--300 GeV energy range, with an
integrated average flux of (93.8 $\pm$ 0.6)$\times$10$^{-8}$ ph cm$^{-2}$
s$^{-1}$, a spectral slope $\alpha$ = 2.16 $\pm$ 0.01 at the reference energy
$E_0$ = 308 MeV, and a curvature parameter around the peak $\beta$ = 0.07
$\pm$ 0.01. The corresponding apparent isotropic $\gamma$-ray luminosity is
(5.1 $\pm$ 0.1) $\times$10$^{48}$ erg s$^{-1}$. As a comparison in the 3FGL catalogue,
covering the period 2008 August 4--2012 July 31, the integrated average flux
is (16.1 $\pm$ 0.5)$\times$10$^{-8}$ ph cm$^{-2}$ s$^{-1}$, and the spectrum
is described by an LP with a spectral slope $\alpha$ = 2.34 $\pm$ 0.03 at the
reference energy $E_0$ = 308 MeV, and a curvature parameter around the peak
$\beta$ = 0.13 $\pm$ 0.02. This indicates a moderate change of the average $\gamma$-ray spectrum during
the period studied here, in which the flux is a factor of approximately six higher than the first four years of LAT operation.  

Fig.~\ref{Fig1} shows the $\gamma$-ray flux (top panel) and spectral parameters (middle panel: spectral slope; bottom panel: curvature parameter) evolution of CTA~102 for the period 2013 January 1--2017 February 9 using an LP model and 30-d time bins. For each time bin, the spectral parameters of both CTA~102 and all sources within 10$^{\circ}$ from it were left free to vary. For the time bins in which the fit results in a TS $<$ 300 for CTA~102, the statistics is not enough for obtaining a detailed characterization of the spectrum with complex
spectral models, therefore we run again the likelihood analysis using an LP model with the curvature parameter fixed to the value obtained integrating over the entire period (i.e $\beta$ = 0.07). The $\gamma$-ray spectrum of
CTA~102 shows a remarkable variability on monthly time-scale, with a spectral slope between
1.88 and 2.97 (the average spectral slope is $\langle \alpha \rangle$ = 2.30 $\pm$ 0.09), and a curvature
parameter between 0.04 and 0.26 (the average curvature parameter is $\langle
\beta \rangle$ = 0.13 $\pm$ 0.04), although for the latter the uncertainties are relatively large.

\begin{figure}
\centering
\includegraphics[width=8.5cm, height=8.5cm, keepaspectratio]{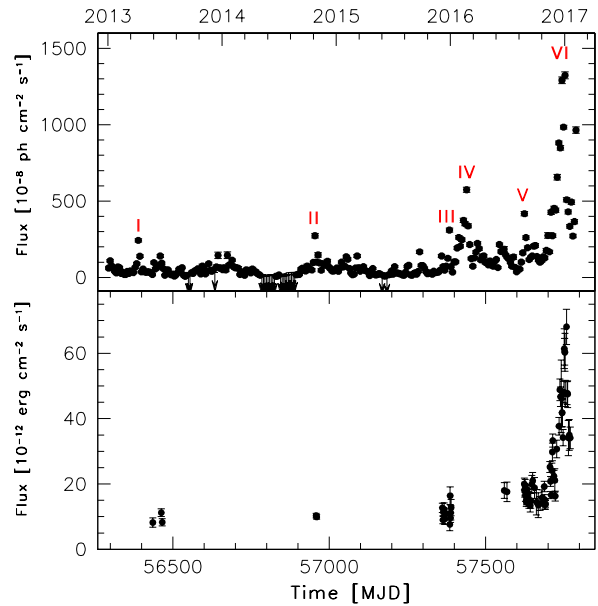}
\caption{Upper panel: Integrated flux light curve of CTA~102 obtained in the 0.1--300 GeV energy range during 2013 January 1--2017 February 9 with five-day time bins. The arrow refers to 2$\sigma$ upper limit on the source flux. Upper limits are computed when TS $<$ 10. Different outbursts are labelled with an identification number in the plot. Bottom panel: X-ray light curve in the 0.3--10 keV energy range obtained by {\em Swift}-XRT (see Section ~\ref{SwiftData} for details).}
\label{Fig2}
\end{figure}

\begin{figure}
\begin{center}
\includegraphics[width=8.5cm, height=8.5cm, keepaspectratio]{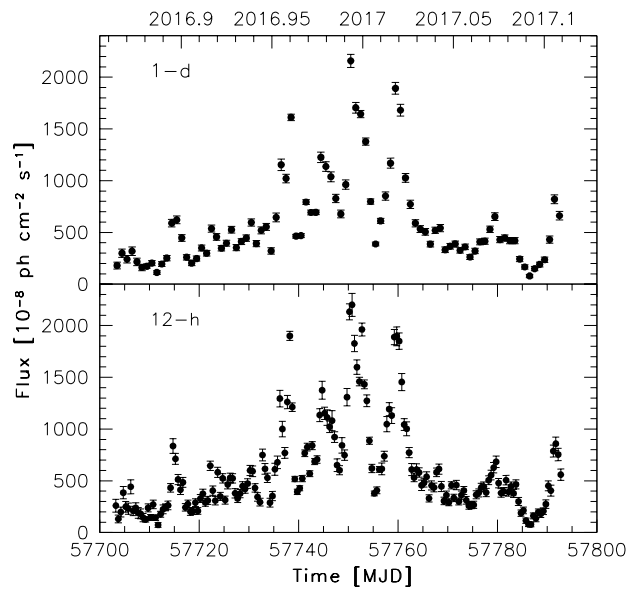}
\caption{\small{Integrated flux light curve of CTA~102 obtained by {\em Fermi}-LAT in the 0.1--300 GeV energy range during 2016 November 11--2017 February 9, with one-day time bins (top panel), and 12-h time bins (bottom panel).}}
\label{Fig3}
\end{center}
\end{figure}

For investigating the $\gamma$-ray variability on different time-scales, we have produced a $\gamma$-ray light curve for the entire period with five-day time bins (Fig.~\ref{Fig2}). For each time bin, the spectral parameters
of CTA~102 and all sources within 10$^{\circ}$ of it were frozen to the values
resulting from the likelihood analysis over the respective monthly
time bin. When TS $<$ 10, 2$\sigma$ upper limits were calculated.  
Six peaks corresponding to periods with fluxes higher than
2$\times$10$^{-6}$ ph cm$^{2}$ s$^{-1}$ were observed in 2013 April
4--8 (MJD 56386--56390; I), 2014 October 21--25 (MJD 56951--56955; II), 2015
December 26--30 (MJD 57382--57386; III), 2016 February
19--23 (MJD 57437--57441; IV), 2016 August 22--26 (MJD 57622--57626; V), and
2016 December 30--2017 January 3 (MJD 57752--57756; VI), with an increase of the flux of a factor between 2.5 and 14 with respect to the average flux estimated during 2013--2017. 

Finally, we have produced a $\gamma$-ray light curve with one-day and 12-h
time bins for the period of high activity, i.e. 2016 November 11--2017
February 9 (MJD 57703--57793), as shown in Fig.~\ref{Fig3}. In the analysis of
the sub-daily light curves, we fixed the flux of the diffuse emission
components at the value obtained by fitting the data over the entire period
analysed in this paper. For each time bin, the spectral parameters of both CTA~102 and all sources within
10$^{\circ}$ of it were frozen to the values resulting from the likelihood
analysis in the monthly time bins. The peak flux of the daily light curve dates 2016 December 28 (MJD 57750), with a flux of (2158 $\pm$ 63)$\times$10$^{-8}$ ph cm$^{-2}$
s$^{-1}$, corresponding to a $\gamma$-ray luminosity (2.2 $\pm$ 0.1) $\times$10$^{50}$ erg s$^{-1}$. A similar peak flux was observed on 12-h time-scales, (2200 $\pm$ 111)$\times$10$^{-8}$ ph cm$^{-2}$ s$^{-1}$ in the second bin of 2016 December 28, corresponding to a $\gamma$-ray luminosity
(2.2 $\pm$ 0.1) $\times$10$^{50}$ erg s$^{-1}$. These values are among the
highest $\gamma$-ray luminosities ever measured for blazars, comparable to what was
observed for 3C 454.3 \citep{abdo11} and S5 0836$+$710 \citep{orienti19}. As a
comparison, in 2012 the $\gamma$-ray flux of CTA~102 reached a peak flux of
$\sim$9 $\times$10$^{-6}$ ph cm$^{-2}$ s$^{-1}$ \citep{lar16}.

The search for variability on very short time-scale in $\gamma$ rays is beyond the
scope of this paper. Rapid variability on time-scale of minutes was observed in 2016 December, with a peak flux of $\sim$3.5 $\times$10$^{-5}$ ph cm$^{-2}$ s$^{-1}$ \citep{gasparyan18, shukla18,meyer19}.

\section{{\em Neil Gehrels Swift Observatory} data: analysis and results}\label{SwiftData}

The {\em Neil Gehrels Swift Observatory} satellite \citep{gehrels04} carried
out 73 observations of CTA~102 between 2013 March 24 (MJD 56436) and
2017 January 18 (MJD 57771). The observations were performed with all three instruments on board: the X-ray Telescope \citep[XRT;][0.2--10.0 keV]{burrows05}, the Ultraviolet/Optical Telescope \citep[UVOT;][170--600 nm]{roming05}, and the Burst Alert Telescope \citep[BAT;][15--150 keV]{barthelmy05}.

The hard X-ray flux of this source turned out to be below the sensitivity of the BAT instrument for such short exposures and therefore the data from this instrument will not be used. Moreover, the source is not present in the {\em Swift} BAT 105-month hard X-ray catalogue \citep{oh18}.

The XRT data were processed with standard procedures (\texttt{xrtpipeline
  v0.13.3}), filtering, and screening criteria by using the \texttt{HEAsoft}
package (v6.22). The data were collected in photon counting mode in all the
observations. The source position in detector coordinates was optimized for each
observation by means of \texttt{XIMAGE}. The source extraction region is
centred on these coordinates.
The source count rate in some observations is higher than 0.5
count s$^{-1}$: these observations were checked for pile-up and a correction
was applied following standard procedures \citep[e.g.,][]{moretti05}. To
correct for pile-up we excluded from the source extraction region the inner
circle of three-pixel radius by considering an annular region with outer radius of
30 pixels (1 pixel $\sim$ 2.36 arcsec). For the other observations source
events were extracted from a circular region with a radius of 20
pixels. Background events were extracted from a circular region with radius of
50 pixels far away from bright sources. Ancillary response files were generated with \texttt{xrtmkarf}, and account for different extraction regions, vignetting
and point spread function corrections. We used the spectral redistribution matrices v014 in the calibration data base maintained by \texttt{HEASARC}\footnote{https://heasarc.gsfc.nasa.gov/docs/heasarc/caldb/}. We fitted the spectrum with an absorbed PL using the
photoelectric absorption model \texttt{tbabs} \citep{wilms00}, with a neutral hydrogen column density fixed to its Galactic value in the source direction \citep[$N_{\rm H}$ = 2.83$\times$10$^{20}$ cm$^{-2}$;][]{kalberla05}. The
results of the fit are reported in Table~\ref{XRTAppendix} and the 0.3--10 keV fluxes are shown in Fig.~\ref{Fig2} in comparison to the $\gamma$-ray light curve obtained by {\em Fermi}-LAT.

\begin{figure}
\includegraphics[width=8.5cm, height=8.5cm, keepaspectratio]{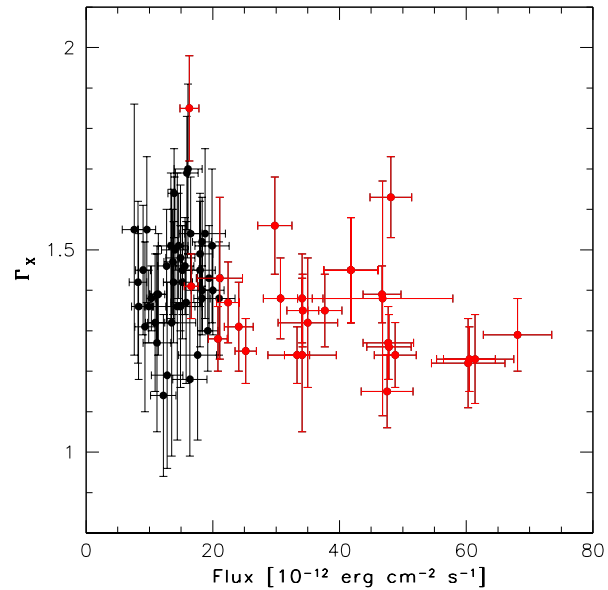}
\caption{{\em Swift}/XRT photon index as a function of the 0.3--10 keV unabsorbed flux. The red points highlight the high-activity period.} 
\label{XRT:hwb}
\end{figure}

The X-ray flux (0.3--10 keV) varied between 7.6$\times$10$^{-13}$ and 68.1$\times$10$^{-13}$ erg cm$^{-2}$ s$^{-1}$ and the photon index between 1.14 and 1.85, with an average value of  $\langle \Gamma_{\rm\,X} \rangle$ = 1.40 $\pm$ 0.15. In Fig.~\ref{XRT:hwb} we plotted the XRT photon index as a function of flux in the 0.3--10 keV energy range. A harder-when-brighter spectral trend has been observed in X-rays in several blazars \citep[e.g.,][]{krawczynski04, dammando11, raiteri12, hayashida15, aleksic15}, although not always present also in the same source \citep[e.g.][]{hayashida12,aleksic17,carnerero17}. This behaviour is usually related to the competition between acceleration and cooling processes acting on relativistic electrons. No spectral hardening with increasing flux is observed either in the entire period or in the high-activity period alone for CTA 102. This suggests that a change in the electron energy distribution is not the main driver of the long-term variability in this energy band. However, at the peak of the X-ray activity the photon index is harder than the average value observed over 2013--2017.

During the {\em Swift} pointings, the UVOT instrument observed CTA~102 in all
its optical ($v$, $b$, and $u$) and UV ($w1$, $m2$, and $w2$) photometric bands
\citep{poole08,breeveld10}. We analysed the data using the \texttt{uvotsource}
task included in the \texttt{HEAsoft} package (v6.22). Source counts were
extracted from a circular region of five-arcsec radius centred on the source,
while background counts were derived from a circular region of 20 arcsec
radius in a nearby source-free region. Observed magnitudes are reported in Table~\ref{UVOTAppendix}. An increase of 4.5--5.5 mag with respect to the
faint state of the source was observed in the UVOT bands, with a range of values: $v=11.44$--16.86, $b=12.56$--17.24, $u=11.78$--16.27, 
$w1=11.33$--16.18, $m2=11.36$--16.17, $w2=11.52$--16.46.

\noindent  Following \citet{raiteri10,raiteri11}, to obtain de-absorbed flux densities we used the count rate to flux density conversion factors $\rm CF$ and amount of Galactic extinction $A_\Lambda$ for each UVOT band that have been obtained by folding the quantities of interest with the source spectrum and effective areas of UVOT filters and are reported in \citet{lar16}. 

Besides correcting flux densities for Galactic extinction, we also subtracted the thermal emission contribution due to the accretion disc and BLR according to the model by \citet{raiteri14}.

\section{Optical-to-radio observations}\label{RadioOptical}

CTA~102 has been monitored by the GLAST-AGILE Support Program (GASP) of the WEBT in the optical, near-infrared and radio bands since 2008. Optical-to-radio GASP-WEBT data collected in 2013--2017 have been presented in \citet{rai17}. That data set is here complemented with data at 15 GHz by the OVRO telescope, at 91 and 103 GHz by ALMA, in the near-IR by the REM telescope, in the optical by the {\em Swift} satellite. The optical photometric observations used in this paper were acquired at the following observatories: Abastumani (Georgia), AstroCamp (Spain), Belogradchik (Bulgaria), Calar Alto (Spain), Campo Imperatore (Italy), Crimean (Russia), Kitt Peak (USA), Lowell (USA; 70 cm, DCT and Perkins telescopes), Lulin (Taiwan), Michael Adrian (Germany), Mt. Maidanak (Uzbekistan), New Mexico Skies (USA), Osaka Kyoiku (Japan), Polakis (USA), Roque de los Muchachos (Spain; Liverpool, NOT and TNG telescopes), ROVOR (USA), Rozhen (Bulgaria; 200 and 50/70 cm telescopes), San Pedro Martir (Mexico), Sirio (Italy), Skinakas (Greece), Steward (USA; Kuiper, Bok, and Super-LOTIS), St. Petersburg (Russia), Teide (Spain), Tien Shan (Kazakhstan), Tijarafe (Spain), Tucson (USA), Valle d'Aosta (Italy), and Vidojevica (Serbia).  

Near-IR data were collected within the WEBT project in the $J$, $H$, and $K$ bands at the Campo Imperatore, Lowell (Perkins) and Teide observatories. These data are here complemented with observations performed by REM \citep{zerbi01, covino04}, a robotic telescope located at the ESO Cerro
La Silla observatory (Chile), in the period 2016 November 15--December 11. All raw near-IR frames obtained with the REM telescope were reduced following standard procedures. Instrumental magnitudes were obtained via aperture photometry and absolute calibration has been performed by means of secondary standard stars in the field reported by
2MASS\footnote{http://www.ipac.caltech.edu/2mass/}. The data presented here
were obtained as Target of Opportunity observations triggered by the
$\gamma$-ray flaring activity of the source (PI: F. D'Ammando). Observed magnitudes are reported in in Table~\ref{REMAppendix}.

Optical and near-IR flux densities were dereddened following the prescriptions of the NASA/IPAC Extragalactic Database\footnote{http://ned.ipac.caltech.edu/} (NED) and corrected for the thermal emission contribution according to the model of \citet{rai17}.

Radio and mm observations were done at the Mets{\"a}hovi Radio Observatory (Finland) at 37 GHz, and at the 30-m IRAM telescope (Spain) and the Sub-millimeter Array (Hawaii, USA) at 230 GHz. We also added the ALMA data collected at 91 and 103 GHz (Band 3) during 2013--2017 and included in the ALMA calibrator source catalogue\footnote{https://almascience.eso.org/alma-data/calibrator-catalogue}.

\noindent As part of an ongoing blazar monitoring program, the OVRO 40-Meter Telescope has observed CTA~102 at 15~GHz
regularly since the beginning of 2009. The OVRO 40-Meter Telescope uses off-axis
dual-beam optics and a cryogenic receiver with 2~GHz equivalent noise
bandwidth centred at 15~GHz. Atmospheric and ground contributions as well as
gain fluctuations are removed with the double switching technique
\citep{readhead89} where the observations are conducted in an ON--ON
fashion so that one of the beams is always pointed on the source. Until 2014 May the two beams were rapidly alternated using a Dicke switch. Since 2014 May, when a new pseudo-correlation receiver replaced the old receiver, a
180~deg phase switch is used. Relative calibration is obtained with a
temperature-stable noise diode to compensate for gain drifts. The primary flux density calibrator is 3C~286, with
an assumed value of 3.44~Jy \citep{baars77}; DR21 is used as secondary
calibrator source. Details of the observation and data reduction schemes are
given in \citet{richards11}. Observations acquired at 15 GHz by OVRO were
used in this paper together with those obtained within the WEBT project. The
radio flux at 15 GHz varied between 2.85 and 3.88 Jy during 2013--2017.

Radio-to-optical light curves collected by WEBT, REM, ALMA, and OVRO will be compared to the $\gamma$-ray light curve obtained by {\em Fermi}-LAT in the Section~\ref{Multifrequency}.

\section{Multifrequency flux variability}\label{Multifrequency}

\begin{table*}
\begin{center}
\caption{Variability amplitude estimated over the period 2013 January 1--2017 February 9 in the different energy bands. Values are corrected for Galactic extinction and the thermal emission contribution. For the minimum and maximum flux density the MJD at which the value is collected is reported in parenthesis.}\label{amplitude}
\begin{tabular}{cccc}
\hline
\multicolumn{1}{c}{Band} &
\multicolumn{1}{c}{Minimum Flux density} &
\multicolumn{1}{c}{Maximum Flux density} &
\multicolumn{1}{c}{Variability amplitude} \\
\multicolumn{1}{c}{} &
\multicolumn{1}{c}{(erg cm$^{-2}$ s$^{-1}$)} &
\multicolumn{1}{c}{(erg cm$^{-2}$ s$^{-1}$)} &
\multicolumn{1}{c}{} \\
\hline
$\gamma$ rays & 5.91$\times$10$^{-13}$ (MJD 57177--57181) & 3.02$\times$10$^{-9}$ (MJD 57752--57756)  & 5110 \\  
X-ray         & 7.60$\times$10$^{-12}$ (MJD 57386) & 6.81$\times$10$^{-11}$ (MJD 57759) &    9 \\ 
W2            & 1.95$\times$10$^{-12}$ (MJD 56958) & 4.92$\times$10$^{-10}$ (MJD 57751) &  253 \\ 
M2            & 3.41$\times$10$^{-12}$ (MJD 56958) & 6.90$\times$10$^{-10}$ (MJD 57752) &  202 \\ 
W1            & 2.55$\times$10$^{-12}$ (MJD 56958) & 5.95$\times$10$^{-10}$ (MJD 57751) &  233 \\ 
U             & 2.50$\times$10$^{-12}$ (MJD 56958) & 3.37$\times$10$^{-10}$ (MJD 57752) &  135 \\ 
B             & 2.98$\times$10$^{-13}$ (MJD 56893) & 7.20$\times$10$^{-10}$ (MJD 57750) & 2416 \\  
V             & 1.97$\times$10$^{-13}$ (MJD 56874) & 7.71$\times$10$^{-10}$ (MJD 57750) & 3920 \\ 
R             & 2.20$\times$10$^{-13}$ (MJD 56785) & 7.76$\times$10$^{-10}$ (MJD 57750) & 3523 \\ 
I             & 3.62$\times$10$^{-13}$ (MJD 56785) & 8.01$\times$10$^{-10}$ (MJD 57750) & 2210 \\  
J             & 6.18$\times$10$^{-13}$ (MJD 56879) & 6.96$\times$10$^{-10}$ (MJD 57752) & 1126 \\ 
H             & 8.59$\times$10$^{-13}$ (MJD 56879) & 7.10$\times$10$^{-10}$ (MJD 57752) & 827  \\   
K             & 1.09$\times$10$^{-12}$ (MJD 56879) & 5.73$\times$10$^{-10}$ (MJD 57751) & 526  \\ 
230 GHz       & 2.25$\times$10$^{-12}$ (MJD 56832) & 1.62$\times$10$^{-11}$ (MJD 57754) &   7  \\
103 GHz       & 1.63$\times$10$^{-12}$ (MJD 56837) & 6.93$\times$10$^{-12}$ (MJD 57649) &  4.3 \\
91  GHz       & 1.51$\times$10$^{-12}$ (MJD 56837) & 5.84$\times$10$^{-12}$ (MJD 57649) &  3.9 \\ 
37 GHz        & 7.92$\times$10$^{-13}$ (MJD 56839) & 1.89$\times$10$^{-12}$ (MJD 57671) &    2 \\ 
15 GHz        & 4.28$\times$10$^{-13}$ (MJD 56651) & 5.99$\times$10$^{-13}$ (MJD 57708) &  1.5 \\
\hline
\end{tabular} 
\end{center}
\end{table*}

Variability studies of radio-to-$\gamma$-ray emission from blazars can provide
important insights into the physics of the jet and the mechanisms at work in
these sources. Flares can be explained e.g. by a shock propagating downstream the
jet and/or by variations of the Doppler factor, which depends on the bulk
Lorentz factor of the relativistic plasma in the jet and on the viewing
angle. During flares, blazars usually show greater variability amplitudes in
the high-energy part of the spectrum than in the low-energy one
\citep[e.g.,][]{wehrle98, raiteri12}. However, some sources increased their brightness by
hundreds of times also in infrared and optical bands. In this context, the 2016--2017 outburst of CTA~102 presented here is one of the best cases to study.

\begin{figure}
\includegraphics[width=8.5cm, height=8.5cm, keepaspectratio]{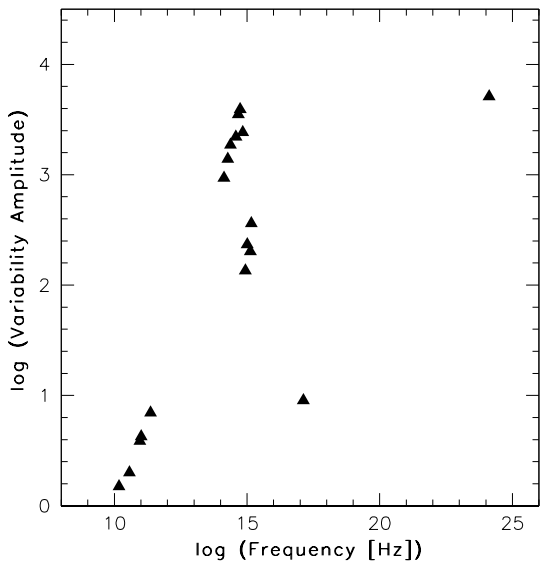}
\caption{Variability amplitude vs frequency in the different energy bands estimated over the period 2013 January 1–-2017 February 9. Values are corrected for Galactic extinction and the thermal emission contribution.}
\label{Variability_Amplitude}
\end{figure}

We quantify the observed variability of the CTA~102 jet emission in the different
energy bands through the variability amplitude, calculated as the ratio of
maximum to minimum flux. Values in the $\gamma$-ray band are based on the light
curve with five-day time bins. Near-infrared-to-UV fluxes are corrected for Galactic extinction. Moreover, the contribution of the thermal emission from the disc, BLR, and torus is removed to properly consider only the jet contribution to the flux. In Table~\ref{amplitude} and Fig.~\ref{Variability_Amplitude}, we report the variability amplitude estimated over the period 2013 January 1--2017 February 9 in the different
bands. The variability amplitude may depend on the sampling of the light curves at the different frequencies. In the case of CTA~102, observations were available in all energy bands at the peak of the flare, making the values obtained representative of the increase of activity of the source.  The variability amplitude shows a rising trend with increasing
frequency in the radio-to-optical range and it declines in the UV. The X-ray
band has a small variability amplitude, in particular if compared to the
$\gamma$-ray one. This can be related to the lower energies of the electrons producing X-rays with respect to those producing $\gamma$ rays.
The similar variability amplitude at 230 GHz and in the X rays may be a hint that they are produced by the same electron population in the same jet region through the synchrotron and IC emission mechanism, respectively, as found e.g.\ for BL Lacertae \citep{raiteri13}.

\begin{figure*}
\includegraphics[width=11cm, height=11cm, keepaspectratio]{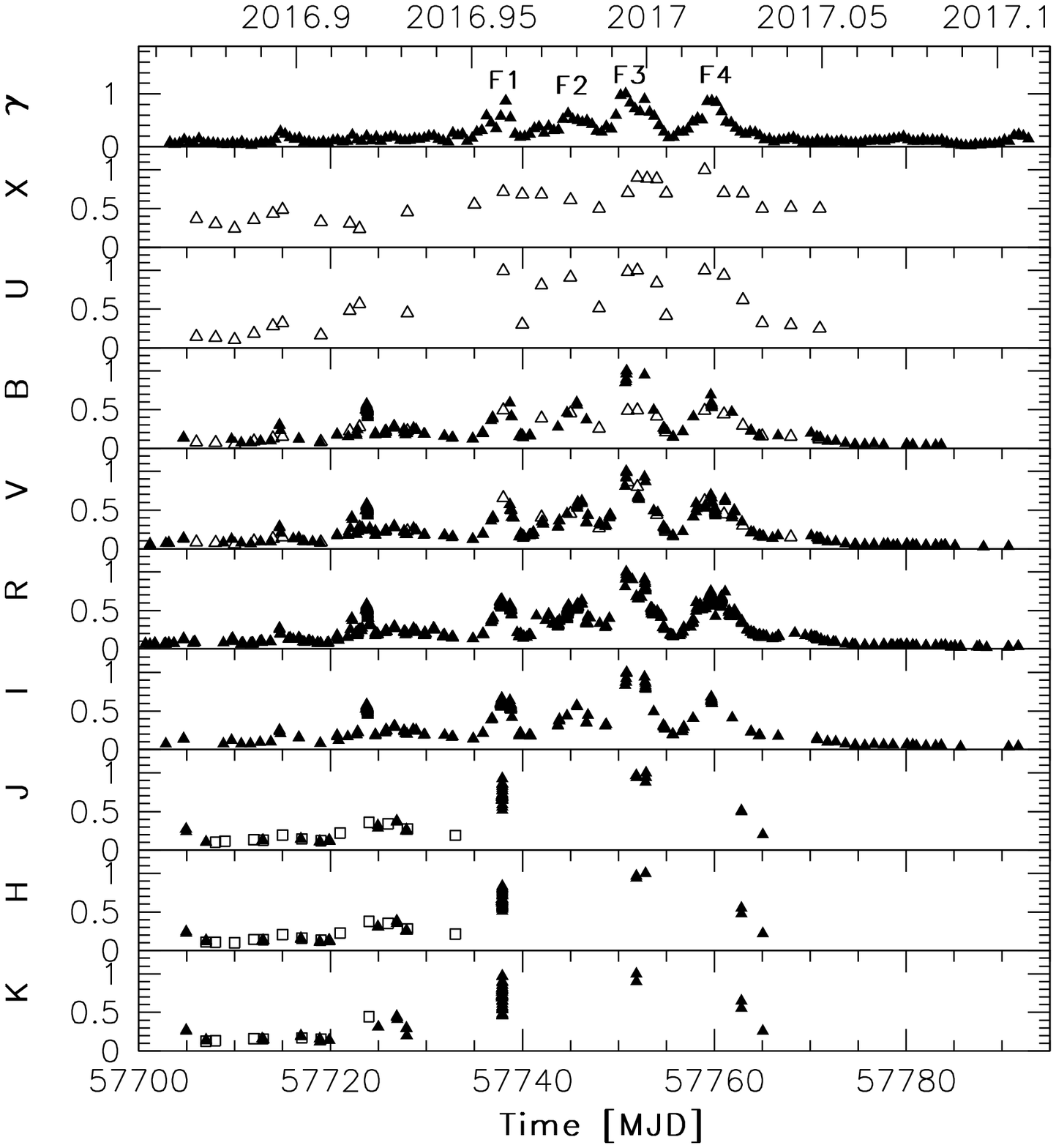}
\caption{Multifrequency light curve normalized to the maximum value observed
  for the period 2016 November 11--2017 February 8 (MJD 57003--57792) in the
  following energy bands (from top to bottom): $\gamma$ rays (100 MeV--300
  GeV), X-rays (0.3--10 keV), $B$, $V$, $R$, $I$, $J$, $H$, and $K$. Filled triangles: WEBT data; open triangles: UVOT data; open squares: REM data. Main $\gamma$-ray outbursts are labelled as F1, F2, F3, and F4 in the top panel.}
\label{BVRIJHK_amplitude}
\end{figure*}

Simultaneous flux variations at low and high energies indicate that their emission comes from the same region of the jet and that the same electrons produce both the synchrotron and IC fluxes \citep[e.g.,][]{fossati08}. However, flaring events in the optical band with no counterpart at high energies were observed in some blazars \citep[e.g.,][]{chatterjee13, dammando13}. Strong variability characterizes the emission over the entire electromagnetic
spectrum of CTA~102 in 2016--2017, in particular during the period 2016 November 11--2017 February 8 (MJD 57003--57792), making this an ideal target to investigate the connection of the flux behaviour observed in $\gamma$ rays with the flux behaviour at lower energies.

\noindent In Fig.~\ref{BVRIJHK_amplitude}, we compare the $\gamma$-ray light
curve obtained by {\em Fermi}-LAT with 12-h time bins during the highest
activity period, 2016 November 11--2017 February 8 (MJD 57003--57792, corresponding to the outburst VI in Fig.~\ref{Fig2}), to the infrared-to-X-ray light curves. All fluxes are normalized to the
maximum value observed in the considered period in order to compare when and
how much the flux increased in the different energy bands. In
the $\gamma$-ray light curve we can see four major outbursts peaked
on 2016 December 15 (MJD 57737; F1), 2016 December 22 (MJD 57744; F2), 2016
December 27 (MJD 57749; F3), and 2017 January 4 (MJD 57757; F4).
The third outburst appears more prominent and shows a larger increase with respect to the others. These four outbursts have
corresponding events in optical, with the peaks observed at the same time. In $J$, $H$, and $K$ bands the sampling is sparse
and only two of the four peaks are observed. A high X-ray flux has been observed in the period that covers the $\gamma$-ray flares F1 and F2, and two X-ray peaks are evident at the time of the $\gamma$-ray flares F3 and F4. 

\begin{figure}
\includegraphics[width=8.5cm, height=8.5cm, keepaspectratio]{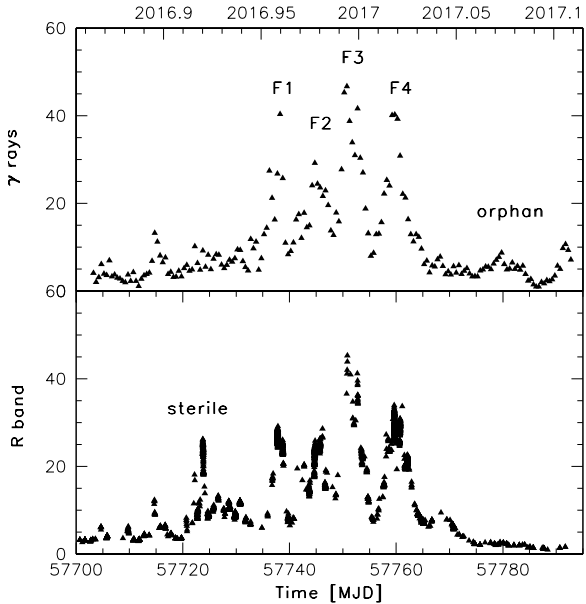}
\caption{Comparison of the {\em Fermi}-LAT $\gamma$-ray light curve with 12-h
  time bins (top panel) and $R$-band light curve (bottom panel) normalized to
  the lowest value observed in the period 2016 November 11--2017 February 8. Main $\gamma$-ray outbursts and ``orphan'' flares are labelled as F1, F2, F3, F4, and orphan in the top panel. The ``sterile'' flare is labelled as sterile in the bottom panel.} 
\label{gamma_optical}
\end{figure}

If we compare the $\gamma$-ray and optical ($R$ band, the best sampled band) fluxes normalized to the
respective lowest values observed during 2016 November 11--2017 February 9
(Fig.~\ref{gamma_optical}), it is evident that the four main flares occurred
at the same time. A similar amplitude has been observed in the two bands, except for the first flare. In particular, the same variability amplitude has been observed during the main peak.
On the other hand, not all the events observed in the optical band have a counterpart in $\gamma$ rays and vice versa. In addition
to the ``sterile" optical flare\footnote{With ``sterile flare'' and ``orphan flare'' we mean a flare observed in the optical band with no counterpart at $\gamma$-ray frequencies, and a flare that is observed at the high energies only, respectively.} occurred around 2016 December 1 (MJD 57723), no significant optical activity corresponds to the increase of $\gamma$-ray activity peaked on 2017 January 24 (MJD 57777) and 2017 February 6 (MJD 57790) (``orphan" flares). This indicates that the interpretation of the source variability must be more complicated than the fair overall optical-$\gamma$ correlation would suggest.

\begin{figure}
\includegraphics[width=8.5cm, height=8.5cm, keepaspectratio]{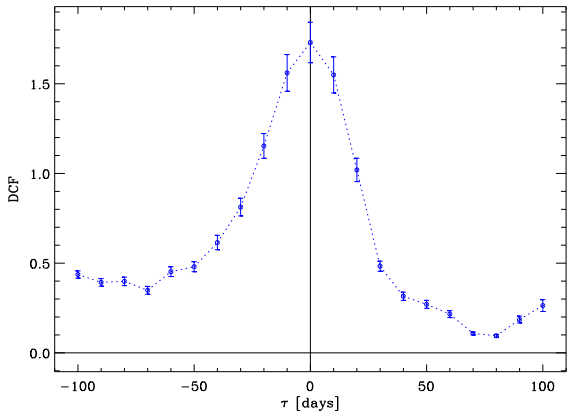}
\caption{DCF between the $\gamma$-ray fluxes obtained with five-day time bins and the $R$-band flux densities with 1 day binning over the whole 2013--2017 period.}
\label{dcf_go_long}
\end{figure}

\begin{figure}
\includegraphics[width=8.5cm, height=8.5cm, keepaspectratio]{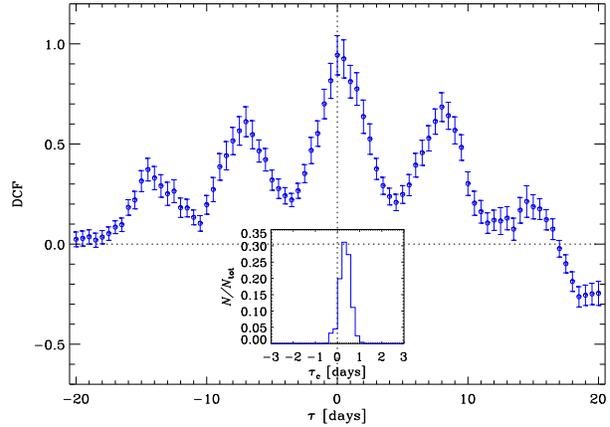}
\caption{DCF between the $\gamma$-ray fluxes obtained with 12-h time bins and the $R$-band flux densities with one-hour binning during the 2016--2017 flaring period. The inset shows the result of cross-correlating 1000 Monte Carlo realizations of the two data sets according to the ``flux redistribution/random subset selection" technique.}
\label{dcf_12h}
\end{figure}

Cross-correlation analysis between flux variations in different bands can allow us to determine whether the emissions come from the same region of the jet or not. We used the discrete correlation function \citep[DCF;][]{edelson88, hufnagel92} to analyze cross-correlations. Correlation produces positive peaks in the DCF and is strong if the peak value approaches or even exceeds one. The DCF between the $\gamma$-ray and the optical ($R$-band) light curves over the whole 2013--2017 period is displayed in Fig.~\ref{dcf_go_long}, showing a main peak compatible with no time lag. When comparing the $\gamma$-ray light curve with 12-h time bin with the optical light curve with one-hour binning in the high-activity period of Fig.~\ref{gamma_optical}, the DCF shows again a main peak compatible with no time lag, with DCF$_p$ = 0.94 (Fig.~\ref{dcf_12h}). This indicates strong correlation between $\gamma$-ray and
optical emission, with no evidence of delay between the flux variations in the two bands, in agreement with the results presented in \citet{lar17}. We determine the uncertainty in this result by performing 1000 Monte Carlo simulations according to the ``flux redistribution/random subset selection'' technique \citep{peterson98,raiteri03}, which tests the importance of sampling and data errors. Among the 1000 simulations, we obtain that 78.3 per cent of simulations ($> 1$ $\sigma$) give a time lag between 0 and 0.6 d, as shown in the inset of Fig.~\ref{dcf_12h}. This is compatible with no delay between optical and $\gamma$-ray emission. The secondary DCF peaks in Fig.~\ref{dcf_12h} are due to the multi-peaked structure of the outburst.

\begin{figure}
\includegraphics[width=8.5cm, height=8.5cm, keepaspectratio]{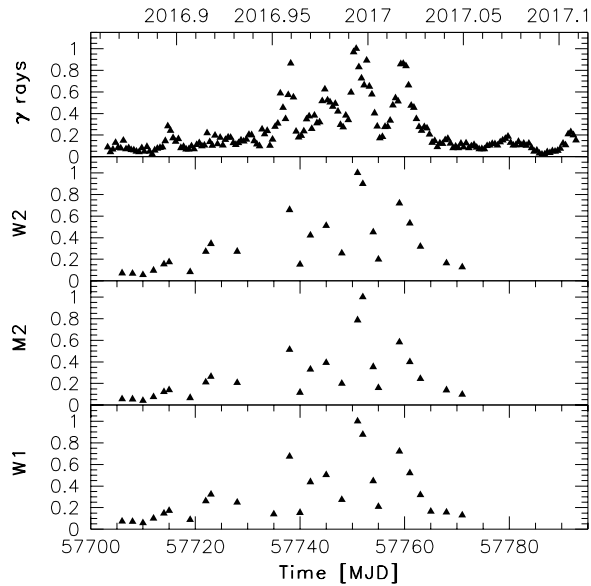}
\caption{Multifrequency light curve normalized to the maximum value observed
  for the period 2016 November 11--2017 February 8 (MJD 57003--57792) in the
  following energy bands (from top to bottom): $\gamma$ rays (100 MeV--300
  GeV), $w2$, $m2$, and $w1$ bands.}
\label{UV_gamma_amplitude}
\end{figure}

Although the UV data collected by {\em Swift}-UVOT are sparser than the optical ones, we can recognize a similar behaviour in the $\gamma$-ray and UV light curves with similar increase of flux during the flaring period when normalized to the maximum value (Fig.~\ref{UV_gamma_amplitude}), even though the variability amplitude is smaller in UV with respect to the $\gamma$-ray band.

\begin{figure}
\includegraphics[width=8.5cm, height=8.5cm, keepaspectratio]{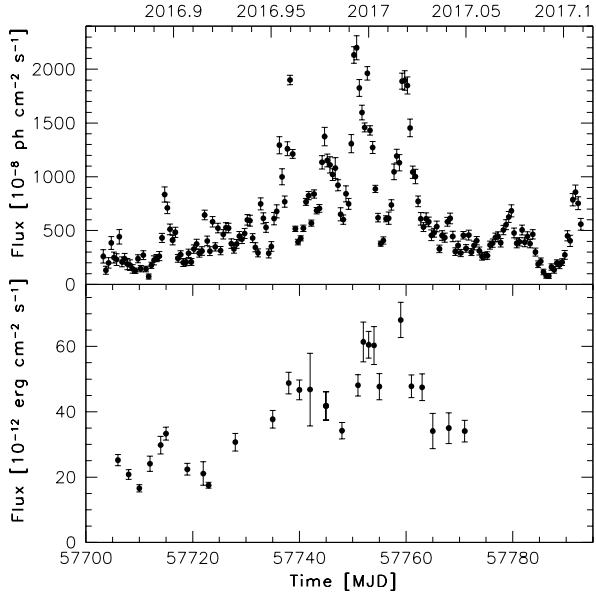}
\caption{{\em Fermi}-LAT $\gamma$-ray light curve with 12-h time bins in the 0.1--300 GeV energy range (top panel) and X-ray light curve in the 0.3--10 keV energy range (bottom panel) in the period 2016 November 11--2017 February 9.} 
\label{gamma_X}
\end{figure}

\begin{figure}
\includegraphics[width=8.5cm, height=8.5cm, keepaspectratio]{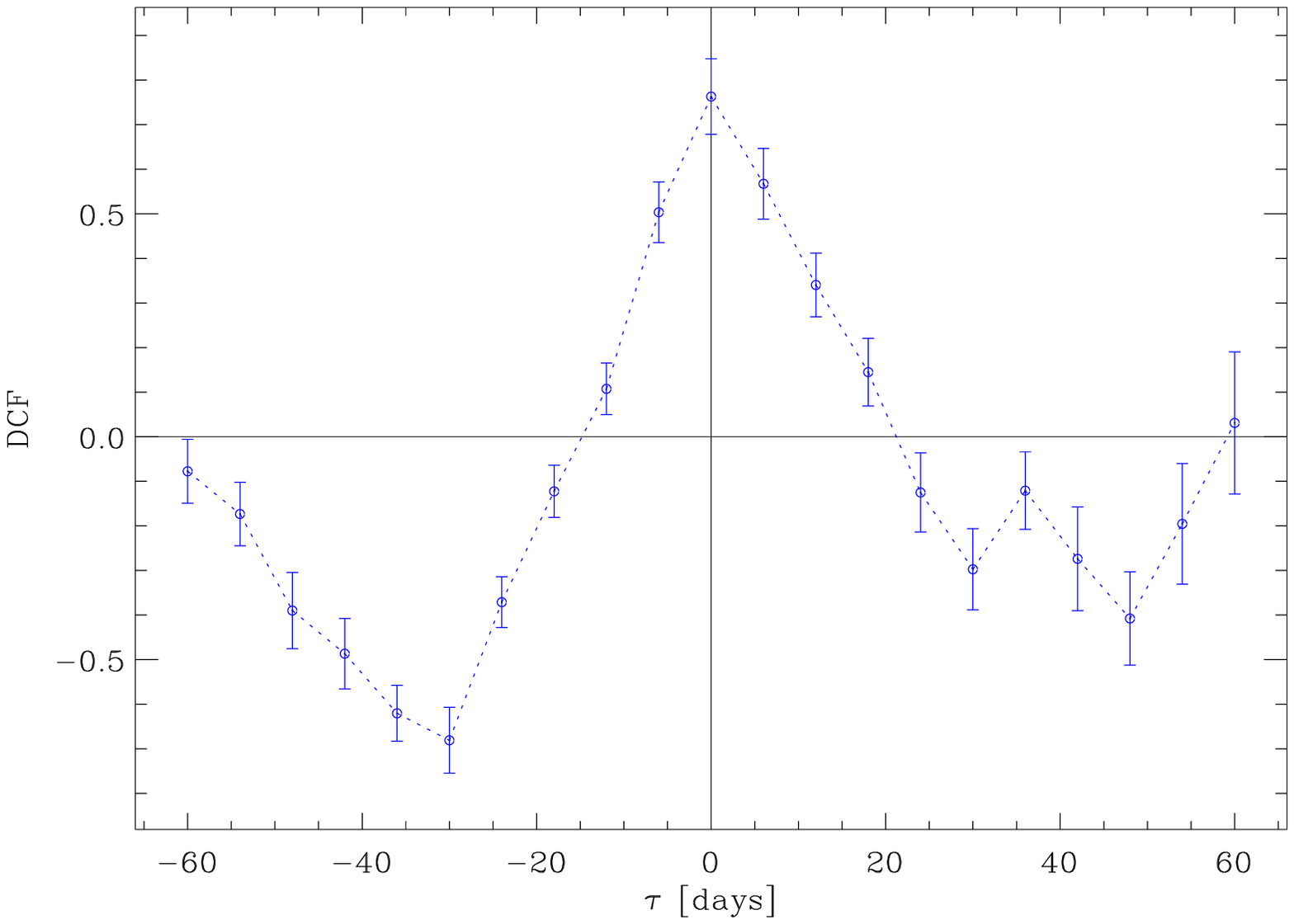}
\caption{DCF between the $\gamma$-ray fluxes obtained with 12-h time bins and the X-ray flux.}
\label{dcf_X}
\end{figure}

The sparse X-ray data in 2013--2015 indicate a low flux, in keeping with the relatively low activity observed also in $\gamma$ rays (see Fig.~\ref{Fig2}). The comparison between the $\gamma$-ray and the X-ray light curves during the high-activity period shows a general agreement, with the $\gamma$-ray peaks corresponding to high X-ray fluxes (Fig.~\ref{gamma_X}). Cross-correlation of the $\gamma$-ray (12-h time bins) and X-ray light curves in the high-activity period shows a good correlation with a time lag compatible to zero within the DCF bin size of six days (Fig.~\ref{dcf_X}). The different sampling of the light curves does not allow a more detailed comparison. However, the X-ray variability amplitude appears much smaller than at $\gamma$ rays. 

\begin{figure}
\includegraphics[width=8.5cm, height=8.5cm, keepaspectratio]{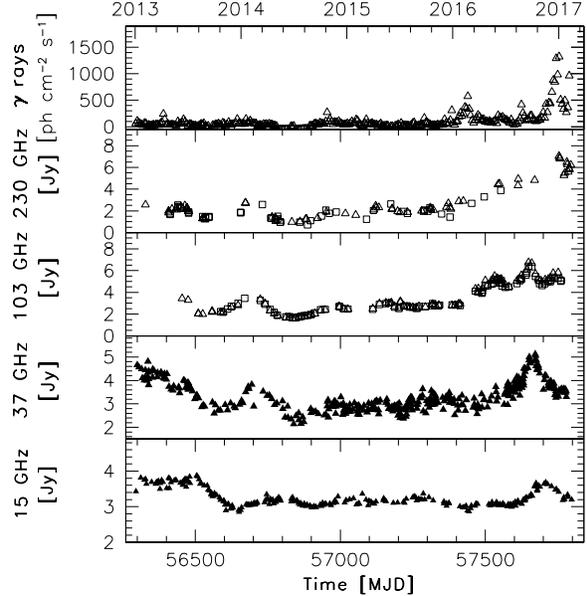}
\caption{Multifrequency light curve for the period 2013 January 1--2017 February 9 in the following energy bands (from top to bottom): $\gamma$ rays (100 MeV--300 GeV; 5-d time bins, in units of 10$^{-8}$ ph cm$^{-2}$ s$^{-1}$; {\em Fermi}-LAT data), 230 GHz (in units of Jy; triangles: SMA data, squares: IRAM data), 91 and 103 GHz (in units of Jy; ALMA data), 37 GHz (in units of Jy; Mets{\"a}hovi data), 15 GHz (in units of Jy; OVRO data).}
\label{radio_lat_ALMA}
\end{figure}

The correlation between the radio-mm fluxes on one side and the optical and $\gamma$-ray fluxes on the other side is rather puzzling. In general, they present a different behavior, but sometimes they share common features. As one can see in Fig.~\ref{radio_lat_ALMA}, the mm data at 230 GHz show a steady flux increase starting from the end of 2015 and culminating with a prominent outburst peaking on 2017 January 1 (MJD 57754). A comparison with the $\gamma$-ray light curve reveals that the end of 2015 also marks the beginning of the activity in this band, with the major peak observed at the same time of the 230 GHz one.

\noindent A steady flux increase starting from the end of 2015 is observed also at 91--103 GHz, is marginally detectable at 37 GHz, and vanishes at 15 GHz. Enhanced activity is present at the beginning of 2014 at 230, 91--103, and 37 GHz, when the light curves at both lower and higher frequencies appear rather flat.

The light curves at 91--103, 37, and 15 GHz present peaks in 2016 October--November. The increase of delay of the radio peaks going to lower frequencies is in agreement with synchrotron self-absorption opacity effects. The decrease of the flux variation amplitudes towards lower frequencies is expected, if the radio emission at lower frequencies comes from more external and extended regions of the jet, in average less aligned to our line of sight.
In and around the same period, the sampling at 230 GHz is poor. The closest events at $\gamma$-ray and optical frequencies date 2016 August (flare V in Fig.~\ref{Fig2}), and, before, there is a stronger $\gamma$-ray flare in 2016 March (flare IV) during a seasonal gap in the optical light curve.

\begin{figure}
\includegraphics[width=8.5cm, height=8.5cm, keepaspectratio]{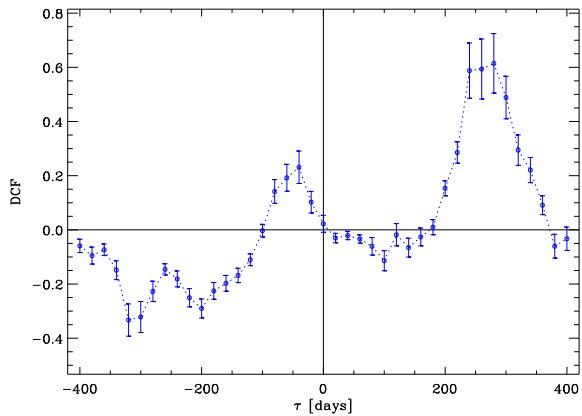}
\caption{DCF between the $\gamma$-ray fluxes obtained with five-day time bins and the 15 GHz flux densities.}
\label{dcf_g15}
\end{figure}

\noindent The DCF between the $\gamma$-ray and 15 GHz fluxes (see Fig.\ \ref{dcf_g15}) shows a small peak at a negative time lag of about two months, indicating radio variations preceding the $\gamma$-ray ones. This signal comes from the cross-match of the 15 GHz flare peaking at the beginning of 2016 November with the $\gamma$-ray acme two months later. A stronger DCF maximum occurs at time lag of about 250 d, indicating radio variations following the $\gamma$-ray ones by about eight to nine months. This suggests that linking the 2016 October--November radio flare observed from 103 to 15 GHz with the 2016 March $\gamma$-ray flare is more likely. Therefore, a delayed radio outburst at 15 GHz can be expected also some months after the 2017 January main flare observed in the $\gamma$-ray, optical, and 230 GHz band.

However, extending the OVRO light curve at 15 GHz up to 2018 July (see Fig.~\ref{OVRO_extended}), a radio outburst is visible only in 2018 February, more than one year after the 2017 $\gamma$-ray, optical and mm peaks. Its peak flux is only a few percent greater than the maximum flux reached in 2016 October--November. On the basis of the delay between $\gamma$-ray emission and radio emission at 15 GHz found here and usually observed in blazar objects \citep[e.g.,][]{pushkarev10, fuhrmann14}, it is unlikely that the 2018 radio outburst at 15 GHz is related to the main flaring activity observed at the beginning of 2017 in $\gamma$-ray, optical, and at 230 GHz. Therefore, we conclude that the $\gamma$-ray flaring activity in 2017 has no (delayed) counterpart in the 15 GHz light curve.

In the framework of the geometrical model, the difference in behaviour between radio and higher energy emission would be ascribed to different viewing angle (with consequent different Doppler boosting of the emission) of the jet regions producing their emission. The extent of misalignment between the emitting jet regions can be inferred from the corresponding light curves, as will be shown in Sect.~\ref{geometry}.  

\section{Multifrequency spectral variability}\label{SpectralVariability}

\begin{figure}
\includegraphics[width=8.5cm, height=8.5cm, keepaspectratio]{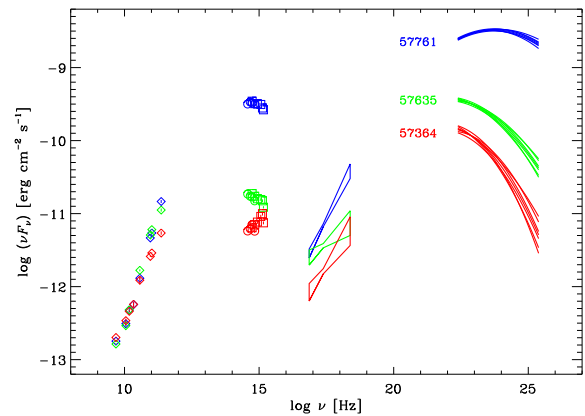}
\caption{Broad-band SED of CTA~102 in three different brightness states labelled with their MJD. In the optical-UV bands, ground-based data are shown as open circles, while UVOT data are displayed with squares. The X-ray spectra are represented by PL fits, while the {\em Fermi}-LAT spectra are plotted according to LP models; in both cases we took the uncertainties on the parameters into account.}
\label{SED}
\end{figure}

Figure~\ref{SED} shows the broad-band SED of CTA~102 in three brightness states
with near-contemporaneous data in the optical, UV, X-ray, and $\gamma$-ray
bands. The optical-to-radio data set presented in \citet{rai17} has been complemented with data at 15 GHz by the OVRO telescope, at 91 and 103 GHz by ALMA, and in the near-IR by the REM telescope, in the optical-to-X-rays by the {\em Swift} satellite, and at $\gamma$ rays by the {\em Fermi} satellite.
In addition to SMA, Mets\"ahovi, ALMA, and OVRO data, radio observations collected by RATAN-600 radio
telescope \citep{mingaliev01} from 1.0 to 21.7 GHz have been considered. Because of the longer variability time-scales of the radio light curves, radio data
are included also if taken within a few days from the reference epoch.

The highest state corresponds to $\rm MJD=57761$ (2017 January 8). Optical
data were acquired at the Tijarafe Observatory and fairly overlap with the
UVOT data. The intermediate state dates $\rm MJD=57635$ (2016 September 4),
and comes from the declining phase of the small flare preceding the big
outburst (flare V in Fig.~\ref{Fig2}). The optical data are from the Mt.\  Maidanak Observatory and are in satisfactorily agreement with the UVOT data. 
For the low state, we chose $\rm MJD=57364$ (2015 December 8), about one year
before the culmination of the big outburst. Optical data are from the
St.\ Petersburg Observatory; the $B$-band point appears a bit fainter than the
corresponding UVOT value, but they agree within errors. For the LAT data the
spectra are extracted on 5-day time-scale as done in Fig.~\ref{Fig2}, except for
the high state in which the daily LAT spectrum is included.   

The intermediate- and the low-state SED become very close in the UV, where the
emission contribution of the big blue bump peaks in coincidence of the
Ly$\alpha\lambda$1216 broad emission lines. The X-ray spectral shape in the
intermediate state is softer than expected, but it is affected by large
errors. Finally, the peak of the IC component shifts at a much higher energy
in the high state, confirming the result found for the synchrotron component
by \citet{rai17}. 

\section{Geometrical model applied to $\gamma$-ray, optical, and radio data}\label{geometry}

In \citet{rai17} we interpreted the long-term variability of the CTA~102 synchrotron flux in terms of variation of the Doppler factor because of changes of the viewing angle of the jet-emitting regions. The intrinsic flux is assumed to be constant on time-scales of months or longer in the rest frame, while flast flares can be due to intrinsic, energetic processes. From the observed multiwavelength light curves we derived how the jet moves, i.e. how the regions emitting at different frequencies align with respect to the line of sight. Support to this twisting jet scenario comes from both observations and theory. Examples of helical jet structures and wobbling motion have been observed with high angular resolution images in the radio band in both extragalactic and Galactic sources \citep[see e.g.,][]{agudo07,agudo12,per12,fro13,bri17,bri18, mil19}.

\noindent In numerical magnetohydrodynamics (MHD) simulations of relativistic jets in 3D, instabilities can develop which distort the jet itself and produce wiggled structures \citep{nak01,mig10}. Moreover, orbital motion in a binary black hole system or a warped accretion disc can lead to jet precession, which modifies the jet orientation with respect to the line of sight \citep{lis18}.

\begin{figure*}
\includegraphics[width=11cm]{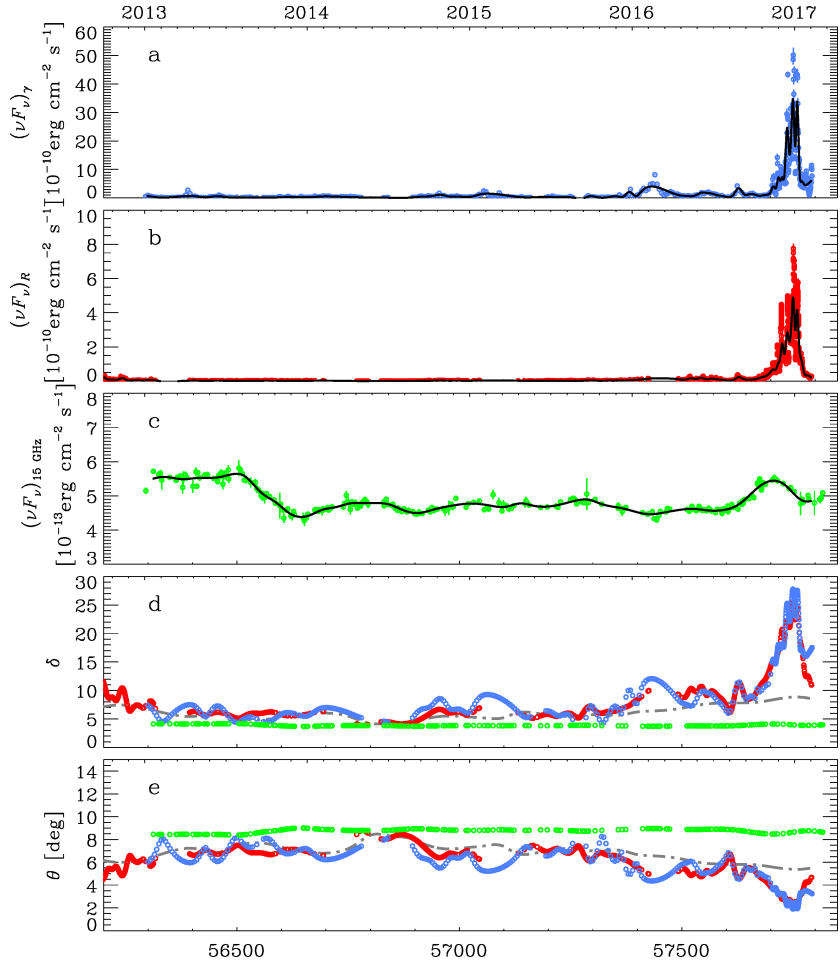}
\caption{Light curves of CTA~102 in the $\gamma$-ray (a), optical (b), and 15 GHz radio (c) bands. The black curves indicate splines interpolations with variable time bin in the optical and $\gamma$-ray bands, and 30-d bins at 15 GHz; they represent the long-term trends. The Doppler factor $\delta$ for the $\gamma$-ray (blue), the optical (red), and the radio (green) bands are plotted in panel (d), while (e) shows the corresponding viewing angles. In panel (d) and (e) the grey dot--dashed curve shows the trend obtained for the 230 GHz data according to \citet{rai17}.}
\label{lc_go}
\end{figure*}

\citet{rai17} performed their analysis on radio--optical data, concentrating on the light curves at 37 and 230 GHz, and in the $R$ band. In the following we investigate the outcome of the proposed geometrical model when applied to both higher and lower frequencies. We examine the $\gamma$-ray and 15 GHz flux variability; the X-ray data are too sparse for a meaningful analysis. We consider the $\gamma$-ray light curve from 2013 January 1 (MJD 56293), with a time bin of five days before 2016 November 11 (MJD 57703) and 12 h after. The $\gamma$-ray light curve is compared to the optical ($R$ band) and radio (15 GHz) light curves in Fig.\ \ref{lc_go}.

As in \citet{rai17}, we modelled the optical long-term trend with a cubic spline interpolation through the data binned with a variable time interval, which shortens as the flux rises: $\Delta t =\Delta t_0 /n$ when $F> F_{\rm min} n^{2+\alpha}$, where the exponent 2 applies to a continuous jet and $\alpha=1.7$ is the spectral index in the $R$ band. We used the same binning to obtain the spline in the $\gamma$-ray band, while for 15 GHz case we adopted a 30-d bin because of the smoother behaviour, as done by \citet{rai17} for the 37 and 230 GHz light curves. 

From the splines we derived the Doppler factors\footnote{The Doppler factor is defined as $\delta=1/[\Gamma \, (1-\beta \cos \theta)]$, where $\Gamma$ is the bulk Lorentz factor, $\beta$ the velocity normalized to the speed of light, and $\theta$ the viewing angle.} $\delta (t)$ displayed in Fig.~\ref{lc_go} as

$$\delta(t)=\delta_{\rm min} \, (F_{\rm spline}/F_{\rm min})^{1/(2+\alpha)}$$

where $F_{\rm spline}$ are the flux densities defined by the spline interpolations, $F_{\rm min}$ are their minimum values, and $\delta_{\rm min}=3.7$ is obtained by assuming a bulk Lorentz factor $\Gamma=20$ and a maximum jet viewing angle $\theta_{\rm max}=9$ deg \citep{rai17}. We adopted a spectral index equal to the optical one for the $\gamma$-ray band and equal to zero for the 15 GHz case.
 
The behaviour of $\delta(t)$ in the optical and $\gamma$-ray bands are in fair general agreement, as expected if the optical and $\gamma$-ray photons are produced in the same jet region. Small differences between the Doppler factor estimated in the optical and $\gamma$-ray bands are justified by the fact that we derived the Doppler factors by interpolating through data affected by errors and collected in bands with different samplings. There is some discordance only in the last part of the considered period, where some $\gamma$-ray activity has no optical counterpart (see discussion in the next section). 

In contrast with the optical and $\gamma$-ray cases, the Doppler factor resulting from the 15 GHz long-term trend is only marginally variable. Actually, in general the Doppler factor and its variability decrease from 230 to 15 GHz, in agreement with the decrease of their flux variability amplitude.

We interpret the time variability of the Doppler factor as due to changes in the jet viewing angle, which implies a twisting jet. The trend of the viewing angle is derived from the definition of the Doppler factor and is shown in Fig.~\ref{lc_go}. The flux maxima correspond to maxima in the Doppler factor and minima in the viewing angle. The plots of $\theta(t)$ suggest that the part of the jet producing the optical and $\gamma$-ray radiation is characterized by intense wiggling, the viewing angle ranging from about nine to two degrees. In contrast, the viewing angle remains more stable as we proceed towards longer wavelengths, i.e. towards outer and larger jet regions. At 15 GHz $\theta(t)$ varies between 8.4 and 9 degrees. Indeed, the viewing angle of the radio-emitting regions can be seen as an average over extended zones of the curved jet. This would also be the reason why the rise of the 230 GHz flux towards the 2016--2017  peak starts much earlier than in the optical and $\gamma$-ray bands. Indeed, while the optical and $\gamma$-ray fluxes reach the minimum viewing angle abruptly, the larger and curved jet section emitting the millimetre flux starts to align with the line of sight even before the optical and $\gamma$-ray-emitting zone. The maximum ``misalignment" between the $\gamma$-optical and 15 GHz emitting regions is reached during the culmination of the 2016--2017 outburst and is about seven degrees.

\begin{figure*}
\includegraphics[width=11cm]{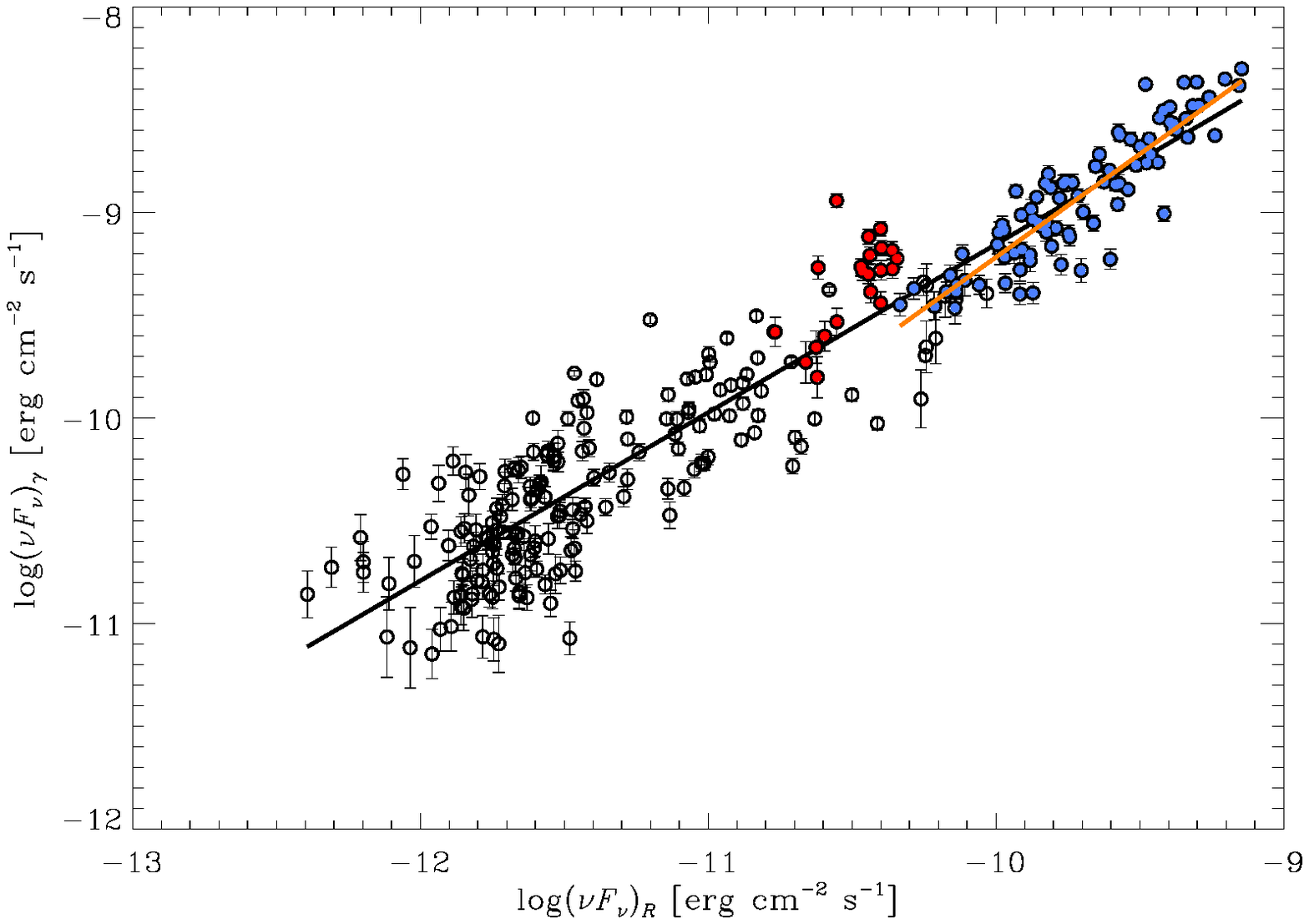}
\caption{The relationship between $\gamma$-ray and optical fluxes. The $\gamma$-ray fluxes have been associated with the average optical fluxes obtained by considering optical data close in time to the $\gamma$-ray optical points. The blue dots refer to the outburst period, the red dots to the period of the $\gamma$-ray ``orphan" flare ($\rm MJD > 57775$). The black and orange solid lines represent linear fits to all the data and to the data in the outbust period, respectively.}
\label{fg_fr}
\end{figure*}

In Fig.~\ref{fg_fr} we show the relationship between $\gamma$-ray and optical fluxes starting from 2013 January 1 (MJD 56293). Each $\gamma$-ray flux is associated with the average of the optical fluxes acquired within 2.5 d from the time of the $\gamma$ data point when the $\gamma$-ray bin is five days, within six hours when the bin is 12 h. The slope of the linear fit is 0.82 and the linear Pearson correlation coefficient is 0.95. 

The geometrical effect is expected to be more evident during the outburst, when we also have better sampling. If we restrict the linear fit to the outburst data in the time interval 2016 November 23--2017 January 22 (MJD 57715--57775\footnote{The range corresponds to a $\gamma$-ray spline flux exceeding $5 \times 10^{-10} \, \rm erg \, cm^{-2} \, s^{-1}$.}), we obtain a linear fit with slope equal to 1.00 and linear Pearson correlation coefficient equal to 0.90. A unity slope is what one expects if both the $\gamma$-ray and optical long-term trends are due to Doppler factor variations \citep{lar16}, so this result supports the idea that the variability during the outburst is mainly caused by changes in $\delta$, as predicted by the geometrical model.

\section{Discussion and conclusions}\label{Summary}

In \citet{rai17} the results of radio-to-optical monitoring of the FSRQ CTA 102 in 2013--2017 were presented and a geometrical model was applied to those data. In this paper, that data set has been complemented with data at 15 GHz by the OVRO telescope, at 91 and 103 GHz by ALMA, in the near-IR by the REM telescope, in the optical-to-X-rays by the {\em Swift} satellite, and at $\gamma$ rays by the {\em Fermi} satellite. These new data have been analysed in view of the above geometrical interpretation.

\noindent Since 2013 April, the source showed significant flux and spectral variability in
$\gamma$ rays on a monthly time-scale, with several periods of high activity. A
general correlation is found between the optical, infrared, and $\gamma$-ray
flux variations, which are consistent with being simultaneous, suggesting that
the observed emission is produced in the same region of the jet.

\noindent The source showed strong activity in the period mid 2016--February 2017 at all frequencies. In particular, an unprecedented $\gamma$-ray flaring activity was observed, reaching  a peak flux of (2158 $\pm$ 63) $\times$10$^{-8}$ ph cm$^{-2}$ s$^{-1}$ in the 0.1--300 GeV energy range on 2016 December 28, corresponding to an apparent isotropic luminosity of (2.2 $\pm$ 0.1)$\times$10$^{50}$ erg s$^{-1}$. 
Comparable values are obtained on 12-h time bin in the same day. The peak luminosity observed for CTA~102 is comparable to the highest values observed in blazars so far. 
Four main outbursts are observed in $\gamma$ rays between 2016 November 11 and 2017 February 9, and corresponding events are
observed in near-IR, optical, and UV with the peaks at the same time. A common
trend was observed also between the X-ray and $\gamma$-ray emission in the
high activity state. No significant harder-when-brighter behaviour was observed in X-rays, indicating that a change in the electron energy distribution is not the main driver of the variability in this band.

{\it The same variability amplitude} was observed in the optical ($R$ band) and $\gamma$ ray bands during the high-activity period. DCF analysis suggests a strong correlation between the flux variations in optical and $\gamma$ rays, with no detectable lag between the emission in the two bands.  However, not all the events observed in the
optical band have a counterpart in $\gamma$ rays and vice versa. The interpretation of these ``orphan flares" is a challenge for all theoretical models aiming to explain blazar variability. They are likely produced in a different emitting region or by a different emission process \citep[e.g., magnetic reconnection;][]{petropoulou16} with respect to the emitting region and mechanism responsible for the long-term behaviour.

\noindent On the other hand, the ``sterile'' optical flare observed around 2016 December 1 (MJD 57723) can be due to the fact that the optical emitting region presents substructures \citep[e.g.,][]{narayan12}, and not all of them produce significant $\gamma$-ray emission. Alternatively, magnitude and direction of a turbulent magnetic field should affect mostly the synchrotron emission, producing an increase of flux in optical not observed in $\gamma$ rays \citep[e.g.,][]{marscher14}. These ``sterile'' and ``orphan'' flares are superposed to the long-term variability well explained by the geometrical model adopted here.

We have investigated the source behaviour at higher ($\gamma$-ray) and lower (15 GHz) frequencies than those analysed in \citet{rai17} by means of their geometrical interpretation of the source variability. This model implies that different emitting regions can change their alignment with respect to the line of sight, leading to a different Doppler boosting and thus enhancement of the observed flux from one region with respect to that from another region. 
 We have derived the trends in time of the Doppler factor and of the viewing angle for the jet regions responsible for the $\gamma$-ray and 15 GHz emission. The trends inferred from the $\gamma$-ray data are in fair agreement with the optical ones and the relationship between the $\gamma$-ray and optical fluxes is linear, as predicted by the model, confirming it. We note that a $\gamma$-ray/optical linear correlation during the outburst may be obtained also with a change of the electron density within a leptonic, one-zone model, where the $\gamma$ rays are produced by an external Compton mechanism. However, the geometrical model also accounts for the continuous time evolution of the multiwavelength flux and not only for single snap-shots of the source behaviour. 

A change in the direction of the jet, which became oriented more closely to our line of sight ($\theta$ $\sim$ 1.2 deg), has been observed with Very Long Baseline Array images at 43 GHz during the $\gamma$-ray flaring activity of CTA~102 in 2012 September--October \citep{casadio15}. These orientation changes can be due to MHD instabilities developing in the jet, precession, or orbital motions in a binary system of supermassive black holes.

Alternative theoretical scenarios have been proposed to explain the flaring activity observed during 2016--2017 in CTA 102. In particular, the multiwavelength flares from 2016 December to 2017 January are explained by  \citet{casadio19} as the interaction between a new jet component and a possible recollimation shock at $\sim$0.1 mas. The variability Doppler factor associated with such interaction (i.e. $\delta_{\rm\,var}$ = 34 $\pm$ 4) is compatible with the values obtained applying our geometrical model (see Fig.~\ref{lc_go}). However, \citet{casadio19} did not try to reproduce the evolution in time of the flaring activity or the long-term behaviour of the multi-wavelength light curves with their model.

\noindent The long-term trend in optical, X-ray, and $\gamma$-ray band has been discussed in a scenario in which the increase of activity is due to the ablation of a gas cloud by the relativistic jet within a hadronic emission model in \citet{zacharias19}. However, the proposed hadronic model results in super-Eddington jet powers at all times. This is a common problem in the application of hadronic models to the SED modelling of blazars \citep[e.g.,][]{boettcher13}. These extreme jet powers obtained in hadronic models are in conflict with the estimates of the jet power based on radio lobes and X-ray cavities \citep[e.g.,][]{merloni07,godfrey13}, and the corresponding accretion rates should imply a very short accretion mode and/or a very small duty cycle in the SMBH evolution compared to estimated lifetimes of active phases of AGN \citep[e.g.,][]{zdziarski15}.  Moreover, the optical and X-ray coverage, both in the flaring and long-term periods, at which the theoretical model proposed in \citet{zacharias19} has been applied appears not adequate to obtain robust results due to long gaps of data and relatively small number of data. The details of the model should be tested with more complete data set like that presented in this paper.
  
We conclude that the observed long-term flux and spectral variability of CTA~102 at both low and high energies can well be explained by an inhomogeneous curved jet where the observed emission at different frequencies is modulated by changes in the orientation of the corresponding emitting regions with respect to our line of sight.

\noindent The main strength of this geometrical model is that it can explain the long-term flux and spectral evolution of CTA~102 in a simple way, with very few assumptions. Its main justification is its ability to explain the contraction of the variability time-scales and the increase of variability amplitude during outburst found by \citet{rai17}. That is the signature of Doppler factor variations which cannot be justified by other models. The analysis performed in this paper, on a wider data set extending to high energies, confirms the model, showing in particular that the relationship between the $\gamma$-ray and optical fluxes is linear, as predicted by the model.

\section*{Acknowledgements}

The data collected by the WEBT collaboration are stored in the WEBT archive at the Osservatorio Astrofisico di Torino - INAF
(http://www.oato.inaf.it/blazars/webt/); for questions regarding their
availability, contact the WEBT President Massimo Villata
(massimo.villata@inaf.it). We acknowledge financial contribution from the agreement ASI-INAF n.\ 2017-14-H.0 and from the contract PRIN-SKA-CTA-INAF 2016.

This research was partially supported by the Bulgarian National Science Fund of the Ministry of Education and Science under grants DN 08-1/2016,
DN 18-13/2017, and KP-06-H28/3 (2018). The Skinakas Observatory is a collaborative project of the University of Crete, the Foundation for
Research and Technology -- Hellas, and the Max-Planck-Institut f\"ur
Extraterrestrische Physik. The Abastumani team acknowledges financial support by
the Shota Rustaveli National Science Foundation under
contract FR/217554/16. The St.Petersburg University team acknowledges support from Russian Science Foundation grant 17-12-01029.
G.Damljanovic and O.Vince gratefully acknowledge the observing grant support from
the Institute of Astronomy and Rozhen National Astronomical Observatory, Bulgarian
Academy of Sciences, via bilateral joint research project "Study of ICRF radio-sources
and fast variable astronomical objects" (head - G.Damljanovic). This work is a part
of the Projects No. 176011 ("Dynamics and Kinematics of Celestial Bodies and
Systems"), No. 176004 ("Stellar Physics"), and No. 176021 ("Visible and Invisible Matter
in Nearby Galaxies: Theory and Observations") supported by the Ministry of Education,
Science and Technological Development of the Republic of Serbia.
JE is indebted to DGAPA (Universidad Nacional Aut\'onoma de M\'exico) for financial support, PAPIIT project IN114917.
The Submillimeter Array is a joint project between the Smithsonian Astrophysical Observatory and the Academia Sinica Institute of Astronomy and Astrophysics and is funded by the Smithsonian Institution and the Academia Sinica.
Data from the Steward Observatory blazar monitoring project were used. This program is supported by NASA/Fermi Guest Investigator grants
NNX12AO93G and NNX15AU81G.
We acknowledge support by Bulgarian National Science Programme "Young Scientists and Postdoctoral Students 2019", Bulgarian National Science
Fund under grant DN18-10/2017 and National RI Roadmap Projects DO1-157/28.08.2018 and DO1-153/28.08.2018 of the Ministry of Education and
Science of the Republic of Bulgaria.
This publication makes use of data obtained at Mets\"ahovi Radio Observatory, operated by Aalto University in Finland. The Astronomical Observatory of the Autonomous Region of the Aosta Valley (OAVdA) is managed by the Fondazione Clément Fillietroz-ONLUS, which is supported by the Regional Government of the Aosta Valley, the Town Municipality of Nus and the "Unité des Communes valdôtaines Mont-Émilius". The research at the OAVdA was partially funded by two "Research and Education" grants from Fondazione CRT. R.R. acknowledges support from CONICYT project Basal AFB-170002. M. Mingaliev and T. Mufakharov acknowledge support through the Russian Government Program of Competitive Growth of Kazan Federal University.

The \textit{Fermi} LAT Collaboration acknowledges generous ongoing support
from a number of agencies and institutes that have supported both the
development and the operation of the LAT as well as scientific data analysis.
These include the National Aeronautics and Space Administration and the
Department of Energy in the United States, the Commissariat \`a l'Energie Atomique
and the Centre National de la Recherche Scientifique / Institut National de Physique
Nucl\'eaire et de Physique des Particules in France, the Agenzia Spaziale Italiana
and the Istituto Nazionale di Fisica Nucleare in Italy, the Ministry of Education,
Culture, Sports, Science and Technology (MEXT), High Energy Accelerator Research
Organization (KEK) and Japan Aerospace Exploration Agency (JAXA) in Japan, and
the K.~A.~Wallenberg Foundation, the Swedish Research Council and the
Swedish National Space Board in Sweden.
 Additional support for science analysis during the operations phase is gratefully
acknowledged from the Istituto Nazionale di Astrofisica in Italy and the Centre
National d'\'Etudes Spatiales in France. This work performed in part under DOE
Contract DE-AC02-76SF00515.

The OVRO 40-m monitoring program is supported in part by NASA grants NNX08AW31G, NNX11A043G and NNX14AQ89G, and NSF grants AST-0808050 and AST-1109911. We thank the {\em Swift} team for making these observations possible, the
duty scientists, and science planners. This research has made use of the NASA/IPAC Extragalactic Database (NED) which is operated by the Jet Propulsion Laboratory, California Institute of Technology, under contract with the National Aeronautics and Space Administration. FD thanks S. Covino for his help with the REM data reduction. This research has made use of data obtained from the high-energy Astrophysics
Science Archive Research Center (HEASARC) provided by NASA's Goddard Space Flight Center.

We regret to mention that K. Kuratov passed away while this paper was under review.

\vspace{5mm}

\noindent $^{1}$INAF - Istituto di Radioastronomia, Via Gobetti 101, I-40129 Bologna,  Italy 
 \\
$^{2}$INAF - Osservatorio Astrofisico di Torino, Via P. Torinese, I- Torino, Italy 
                                                                                \\
$^{ 3}$Instituto de Astrofisica de Canarias (IAC), La Laguna, E-38200 Tenerife, Spain                                                                        \\
$^{ 4}$Departamento de Astrofisica, Universidad de La Laguna, La Laguna, E-38205 Tenerife, Spain                                                             \\
$^{ 5}$Instituto de Astrof\'{\i}sica de Andaluc\'{\i}a (CSIC), E-18080 Granada, Spain                                                                        \\
$^{ 6}$Pulkovo Observatory, 196140 St.\ Petersburg, Russia                                                                                                   \\
$^{ 7}$Institute of Astronomy and NAO, Bulgarian Academy of Sciences, 1784 Sofia, Bulgaria                                                                   \\
$^{ 8}$Crimean Astrophysical Observatory RAS, P/O Nauchny, 298409, Russia                                                                                    \\
$^{ 9}$Instituto de Astronom\'ia, Universidad Nacional Aut\'onoma de M\'exico, 04510 Ciudad de M\'exico, M\'exico                                                                      \\
$^{ 10}$INAF, TNG Fundaci\'on Galileo Galilei, E-38712 La Palma, Spain                                                                                        \\
$^{11}$Department of Astronomy, Faculty of Physics, University of Sofia, BG-1164 Sofia, Bulgaria                                                             \\
$^{12}$Osservatorio Astronomico della Regione Autonoma Valle d'Aosta, I-11020, Nus, Italy                                                                     \\
$^{13}$EPT Observatories, Tijarafe, E-38780 La Palma, Spain                                                                                                  \\
$^{14}$Max-Planck-Institut f\"ur Radioastronomie, D--53121, Bonn, Germany                                                                                    \\
$^{15}$School of Physics \& Astronomy, University of Southampton, Southampton, SO17 1BJ, UK                                                                  \\
$^{16}$Graduate Institute of Astronomy, National Central University, Jhongli City, Taoyuan County 32001, Taiwan        
          \\
$^{17}$Astronomical Observatory, 11060 Belgrade, Serbia                                                                                                      \\
$^{18}$INAF, Osservatorio Astronomico di Roma, I-00040 Monte Porzio Catone, Italy                                                                            \\
$^{19}$Ulugh Beg Astronomical Institute, Maidanak Observatory, Tashkent, 100052, Uzbekistan                                                                  \\
$^{20}$Astronomical Institute, St.\ Petersburg State University, 198504 St.\ Petersburg, Russia                                                              \\
$^{21}$Center for Astrophysics \textbar\, Harvard \& Smithsonian, 60 Garden Street, Cambridge, MA 02138 USA
          \\
$^{22}$Astrophysics Research Institute, Liverpool John Moores University, Liverpool L3 5RF, UK                                                               \\
$^{23}$School of Cosmic Physics, Dublin Institute For Advanced Studies, Ireland                                                                              \\
$^{24}$Institute for Astrophysical Research, Boston University, Boston, MA 02215, USA                                                                        \\
$^{25}$NNLOT, Al-Farabi Kazakh National University, 050040 Almaty, Kazakhstan                                                                                       \\
$^{26}$Fesenkov Astrophysical Institute, Almaty, Kazakhstan                                                                                                  \\
$^{27}$Abastumani Observatory, Mt. Kanobili, 0301 Abastumani, Georgia                                                                                        \\
$^{28}$Engelhardt Astronomical Observatory, Kazan Federal University, 422526 Tatarstan, Russia
          \\
$^{29}$Landessternwarte, Zentrum für Astronomie der Universität Heidelberg, 69117 Heidelberg, Germany   
          \\
$^{30}$Center for Astrophysics, Guangzhou University, Guangzhou, 510006, China                                                                               \\
$^{31}$Aalto University Mets\"ahovi Radio Observatory, FI-02540 Kylm\"al\"a, Finland                                                                         \\
$^{32}$Aalto University Dept of Electronics and Nanoengineering, FI-00076 Aalto, Finland                                                           
          \\                                                                                                
$^{33}$Astronomical Institute, Osaka Kyoiku University, Osaka, 582-8582, Japan                   \\
$^{34}$UCD School of Physics, University College Dublin, Dublin 4, Ireland                                                                                   \\
$^{35}$Department of Physics and Astronomy, Brigham Young University, Provo, UT 84602, USA                                                                   \\
$^{36}$Michael Adrian Observatorium, Astronomie Stiftung Trebur, 65468 Trebur, Germany                                                                       \\
$^{37}$University of Applied Sciences, Technische Hochschule Mittelhessen, 61169 Friedberg, Germany                                                          \\
$^{38}$Command Module Observatory, 121 W. Alameda Dr., Tempe, AZ 85282, USA                                                                                                            \\
$^{39}$Nordic Optical Telescope, E-38700 Santa Cruz de La Palma, Santa Cruz de Tenerife, Spain                                                               \\
$^{40}$Osservatorio Astronomico Sirio, I-70013 Castellana Grotte, Italy                                                                                      \\
$^{41}$LESIA, Observatoire de Paris, Universit\'e PSL, CNRS, Sorbonne Universit\'e, Univ. Paris Diderot, Sorbonne Paris Cit\'e, 5 place Jules
Janssen, 92195 Meudon, France
          \\
$^{42}$Department of Physics, University of Colorado Denver, CO, 80217-3364 USA                                                                              \\
$^{43}$Lowell Observatory, Flagstaff, AZ 85751, USA                                                                                                          \\
$^{44}$Steward Observatory, University of Arizona, Tucson, AZ 85721, USA                                                                                     \\
$^{45}$Instituto de Radio Astronom\'ia Milim\'etrica, E-18012 Granada, Spain  
          \\
$^{46}$Finnish Center for Astronomy with ESO (FINCA), University of Turku, FI-20014, Turku, Finland 
          \\     
$^{47}$Aalto University Metsähovi Radio Observatory, Metsähovintie 114, 02540 Kylm\"al\"a, Finland 
          \\
$^{48}$Owens Valley Radio Observatory, California Institute of Technology, Pasadena, CA 91125, USA 
          \\
$^{49}$Departamento de Astronomía, Universidad de Chile, Camino El Observatorio 1515, Las Condes, Santiago, Chile 
          \\
$^{50}$Departamento de Astronomia, Universidad de Concepcion, Concepcion, Chile 
          \\
$^{51}$Shanghai Astronomical Observatory, Chinese Academy of Sciences, Shanghai 200030, China
          \\
$^{52}$Kazan Federal University, 18 Kremlyovskaya St., Kazan 420044, Russia
          \\
$^{53}$Special Astrophysical Observatory of RAS, Nizhnij Arkhyz 369167, Russia
          \\

\clearpage

\appendix

\onecolumn

\section{Online Tables and Figures} \label{Appendix}

\setcounter{table}{0}
\begin{table*}
  \caption{Log and fitting results of {\em Swift}-XRT observations of CTA~102 using a PL model with $N_{\rm H}$ fixed to Galactic absorption. Fluxes are corrected for the Galactic absorption.} 
\label{XRTAppendix}
\begin{center}
\begin{tabular}{ccccc}
\hline
\multicolumn{1}{c}{Date (UT)} &
\multicolumn{1}{c}{MJD} &
\multicolumn{1}{c}{Net exposure time} &
\multicolumn{1}{c}{Flux 0.3--10 keV}  &
\multicolumn{1}{c}{Photon index}  \\
\multicolumn{1}{c}{} &
\multicolumn{1}{c}{} &
\multicolumn{1}{c}{(s)} &
\multicolumn{1}{c}{($\times$10$^{-13}$ erg cm$^{-2}$ s$^{-1}$)} &
\multicolumn{1}{c}{($\Gamma_{\rm\,X}$)} \\
\hline
2013-May-24 & 56436 &  1041 &  $8.2  \pm 1.4$ & $1.42 \pm 0.20$ \\ 
2013-Jun-20 & 56463 &  3019 &  $11.2 \pm 1.2$ & $1.39 \pm 0.12$ \\ 
2013-Jun-23 & 56466 &  2974 &  $8.3  \pm 1.1$ & $1.36 \pm 0.18$ \\
2014-Oct-28 & 56958 &  4942 &  $10.3 \pm 0.7$ & $1.38 \pm 0.08$ \\ 
2014-Oct-30 & 56960 &  4942 &  $9.9  \pm 0.7$ & $1.36 \pm 0.09$ \\ 
2015-Dec-06 & 57362 &  1940 &  $12.7 \pm 1.4$ & $1.46 \pm 0.14$ \\ 
2015-Dec-08 & 57364 &  1793 &  $9.0  \pm 1.2$ & $1.45 \pm 0.16$ \\
2015-Dec-09 & 57365 &  1970 &  $11.4 \pm 1.3$ & $1.39 \pm 0.15$ \\
2015-Dec-12 & 57368 &  1616 &  $12.2 \pm 2.0$ & $1.14 \pm 0.20$ \\ 
2015-Dec-14 & 57370 &  1603 &  $9.3  \pm 1.6$ & $1.31 \pm 0.21$ \\
2015-Dec-18 & 57374 &  1766 &  $10.9 \pm 1.5$ & $1.32 \pm 0.17$ \\ 
2015-Dec-30 & 57386 &   762 &  $7.6  \pm 1.9$ & $1.55 \pm 0.31$ \\ 
2015-Dec-31 & 57387 &   979 &  $16.4 \pm 2.7$ & $1.18 \pm 0.19$ \\ 
2016-Jan-01 & 57388 &  1986 &  $9.6  \pm 1.4$ & $1.55 \pm 0.18$ \\ 
2016-Jan-02 & 57389 &   817 &  $11.2 \pm 2.2$ & $1.27 \pm 0.22$ \\ 
2016-Jan-03 & 57390 &   111 &  $12.8 \pm 2.5$ & $1.19 \pm 0.23$ \\ 
2016-Jun-21 & 57560 &   976 &  $18.0 \pm 2.4$ & $1.45 \pm 0.19$ \\ 
2016-Jun-30 & 57569 &   844 &  $17.6 \pm 3.0$ & $1.24 \pm 0.21$ \\ 
2016-Aug-24 & 57624 &  1621 &  $20.0 \pm 1.8$ & $1.40 \pm 0.11$ \\ 
2016-Aug-25 & 57625 &  1668 &  $18.0 \pm 2.0$ & $1.49 \pm 0.13$ \\ 
2016-Aug-26 & 57626 &  1860 &  $18.3 \pm 1.7$ & $1.40 \pm 0.11$ \\ 
2016-Aug-27 & 57627 &  1446 &  $19.4 \pm 2.0$ & $1.43 \pm 0.13$ \\ 
2016-Aug-28 & 57628 &  2128 &  $16.5 \pm 1.8$ & $1.54 \pm 0.14$ \\ 
2016-Aug-29 & 57629 &  2797 &  $18.3 \pm 1.3$ & $1.38 \pm 0.08$ \\ 
2016-Aug-30 & 57630 &  1558 &  $14.6 \pm 1.5$ & $1.51 \pm 0.13$ \\ 
2016-Aug-31 & 57631 &  2133 &  $15.2 \pm 1.6$ & $1.42 \pm 0.14$ \\ 
2016-Sep-02 & 57633 &  1963 &  $15.6 \pm 1.4$ & $1.46 \pm 0.12$ \\ 
2016-Sep-02 & 57633 &  1963 &  $18.3 \pm 1.5$ & $1.52 \pm 0.11$ \\ 
2016-Sep-03 & 57634 &   959 &  $16.1 \pm 2.2$ & $1.70 \pm 0.21$ \\ 
2016-Sep-04 & 57635 &   854 &  $15.9 \pm 1.8$ & $1.69 \pm 0.14$ \\ 
2016-Sep-08 & 57639 &   971 &  $14.9 \pm 2.3$ & $1.48 \pm 0.18$ \\ 
2016-Sep-12 & 57643 &   892 &  $13.4 \pm 2.0$ & $1.51 \pm 0.18$ \\
2016-Sep-14 & 57645 &  1076 &  $14.9 \pm 2.3$ & $1.36 \pm 0.20$ \\ 
2016-Sep-17 & 57648 &   874 &  $19.9 \pm 2.7$ & $1.51 \pm 0.19$ \\ 
2016-Sep-20 & 57651 &   991 &  $21.0 \pm 2.5$ & $1.38 \pm 0.14$ \\ 
2016-Sep-26 & 57657 &   772 &  $18.8 \pm 3.2$ & $1.54 \pm 0.21$ \\ 
2016-Oct-02 & 57663 &   332 &  $14.4 \pm 3.9$ & $1.36 \pm 0.33$ \\ 
2016-Oct-08 & 57669 &   587 &  $13.5 \pm 3.8$ & $1.32 \pm 0.33$ \\ 
2016-Oct-14 & 57675 &   946 &  $14.0 \pm 2.1$ & $1.50 \pm 0.18$ \\ 
2016-Oct-20 & 57681 &   969 &  $15.8 \pm 2.4$ & $1.37 \pm 0.20$ \\ 
2016-Oct-27 & 57688 &  1953 &  $19.2 \pm 1.6$ & $1.30 \pm 0.10$ \\ 
2016-Oct-28 & 57689 &  1698 &  $15.2 \pm 1.5$ & $1.45 \pm 0.11$ \\ 
2016-Oct-29 & 57690 &  1723 &  $13.7 \pm 1.4$ & $1.47 \pm 0.13$ \\ 
2016-Oct-30 & 57691 &  1651 &  $13.8 \pm 1.6$ & $1.42 \pm 0.15$ \\ 
2016-Oct-31 & 57692 &  2100 &  $13.9 \pm 1.0$ & $1.64 \pm 0.11$ \\ 
2016-Nov-14 & 57706 &  2974 &  $25.2 \pm 1.7$ & $1.25 \pm 0.08$ \\ 
2016-Nov-16 & 57708 &  2742 &  $20.8 \pm 1.5$ & $1.28 \pm 0.08$ \\ 
2016-Nov-18 & 57710 &  3119 &  $16.6 \pm 1.1$ & $1.41 \pm 0.08$ \\ 
2016-Nov-20 & 57712 &  2420 &  $24.1 \pm 2.3$ & $1.31 \pm 0.11$ \\ 
2016-Nov-22 & 57714 &  1683 &  $29.8 \pm 2.7$ & $1.56 \pm 0.12$ \\ 
2016-Nov-23 & 57715 &  2919 &  $33.3 \pm 2.0$ & $1.24 \pm 0.07$ \\ 
2016-Nov-27 & 57719 &  1933 &  $22.4 \pm 1.8$ & $1.37 \pm 0.10$ \\ 
2016-Nov-30 & 57722 &   380 &  $21.1 \pm 3.6$ & $1.43 \pm 0.20$ \\ 
2016-Dec-01 & 57723 &  1325 &  $16.3 \pm 1.5$ & $1.85 \pm 0.13$ \\ 
2016-Dec-06 & 57728 &  1938 &  $30.7 \pm 2.7$ & $1.38 \pm 0.10$ \\ 
2016-Dec-13 & 57735 &  2647 &  $37.7 \pm 2.7$ & $1.35 \pm 0.09$ \\ 
2016-Dec-16 & 57738 &  2432 &  $48.8 \pm 3.3$ & $1.24 \pm 0.08$ \\ 
2016-Dec-18 & 57740 &  2398 &  $46.7 \pm 3.0$ & $1.39 \pm 0.07$ \\ 
2016-Dec-20 & 57742 &   804 &  $46.8 \pm 11.1$& $1.38 \pm 0.29$ \\ 
2016-Dec-23 & 57745 &  1978 &  $41.8 \pm 4.3$ & $1.45 \pm 0.13$ \\ 
2016-Dec-26 & 57748 &  1667 &  $34.2 \pm 2.5$ & $1.35 \pm 0.09$ \\ 
2016-Dec-29 & 57751 &  1808 &  $48.1 \pm 3.3$ & $1.63 \pm 0.10$ \\ 
\hline
\end{tabular}
\end{center}
\end{table*}

\addtocounter{table}{-1}
\begin{table*}
\begin{center}
\begin{tabular}{ccccc}
\hline
\multicolumn{1}{c}{Date (UT)} &
\multicolumn{1}{c}{MJD} &
\multicolumn{1}{c}{Net exposure time} &
\multicolumn{1}{c}{Flux 0.3--10 keV}  &
\multicolumn{1}{c}{Photon index}  \\
\multicolumn{1}{c}{} &
\multicolumn{1}{c}{} &
\multicolumn{1}{c}{(s)} &
\multicolumn{1}{c}{($\times$10$^{-13}$ erg cm$^{-2}$ s$^{-1}$)} &
\multicolumn{1}{c}{($\Gamma_{\rm\,X}$)} \\
\hline
2016-Dec-30 & 57752 &  1474 &  $61.4 \pm 6.1$ & $1.23 \pm 0.11$ \\ 
2016-Dec-31 & 57753 &  2040 &  $60.5 \pm 4.1$ & $1.23 \pm 0.08$ \\ 
2017-Jan-01 & 57754 &  1484 &  $60.3 \pm 5.8$ & $1.22 \pm 0.11$ \\
2017-Jan-02 & 57755 &  1552 &  $47.7 \pm 4.0$ & $1.27 \pm 0.09$ \\ 
2017-Jan-06 & 57759 &  2452 &  $68.1 \pm 5.4$ & $1.29 \pm 0.09$ \\ 
2017-Jan-08 & 57761 &  2480 &  $47.8 \pm 3.5$ & $1.26 \pm 0.08$ \\ 
2017-Jan-10 & 57763 &  2480 &  $47.5 \pm 4.1$ & $1.15 \pm 0.09$ \\ 
2017-Jan-12 & 57765 &   517 &  $34.1 \pm 5.4$ & $1.24 \pm 0.19$ \\ 
2017-Jan-15 & 57768 &   991 &  $35.0 \pm 4.7$ & $1.32 \pm 0.16$ \\ 
2017-Jan-18 & 57771 &  1746 &  $34.1 \pm 3.3$ & $1.38 \pm 0.11$  \\ 
\hline
\end{tabular}
\end{center}
\end{table*}

\newpage

\setcounter{table}{1}
\begin{table*}
\caption{Observed magnitude of CTA~102 obtained by {\em Swift}-UVOT.}
\label{UVOTAppendix}
\begin{center}
\begin{tabular}{cccccccc}
\hline
\multicolumn{1}{c}{Date (UT)} &
\multicolumn{1}{c}{MJD}       &
\multicolumn{1}{c}{$v$}         &
\multicolumn{1}{c}{$b$}         &
\multicolumn{1}{c}{$u$}         &
\multicolumn{1}{c}{$w1$}        &
\multicolumn{1}{c}{$m2$}        &
\multicolumn{1}{c}{$w2$}        \\
\hline
2013-May-24 & 56436 &  $16.86 \pm 0.12$ & $17.18 \pm 0.09$ & $16.17 \pm 0.07$ & $16.04 \pm 0.08$ & $16.09 \pm 0.06$ & $ 16.28 \pm 0.07$ \\ 
2013-Jun-20 & 56463 &  --  & -- & -- & $16.03 \pm  0.06$ & -- & $16.06 \pm 0.06$ \\ 
2013-Jun-23 & 56466 &  --  & -- & -- & -- & $16.06 \pm 0.06$ & --  \\ 
2014-Oct-28 & 56958 &  $16.77 \pm 0.11$ & $17.11 \pm 0.09$ & $16.27 \pm 0.08$ & $16.18 \pm 0.09$ & $16.17 \pm 0.06$ & $16.46 \pm 0.08$ \\ 
2014-Oct-30 & 56960 &  $16.55 \pm 0.08$ & $17.01 \pm 0.07$ & $16.03 \pm 0.06$ & $16.02 \pm 0.07$ & $15.99 \pm 0.08$ & $16.29 \pm 0.07$ \\ 
2015-Dec-06 & 57362 &  $15.82 \pm 0.08$ & $16.46 \pm 0.08$ & $15.49 \pm 0.07$ & $15.45 \pm 0.08$ & $15.61 \pm 0.09$ & $15.81 \pm 0.09$ \\
2015-Dec-08 & 57364 &  $16.49 \pm 0.11$ & $16.83 \pm 0.09$ & $15.90 \pm 0.08$ & $15.73 \pm 0.09$ & $15.71 \pm 0.09$ & $16.07 \pm 0.08$ \\ 
2015-Dec-09 & 57365 &  $16.61 \pm 0.14$ & $16.99 \pm 0.11$ & $15.94 \pm 0.09$ & $15.80 \pm 0.10$ & $15.98 \pm 0.11$ & $16.13 \pm 0.09$ \\ 
2015-Dec-12 & 57368 &  $16.78 \pm 0.13$ & $17.04 \pm 0.10$ & $15.95 \pm 0.08$ & $15.93 \pm 0.09$ & $15.91 \pm 0.10$ & $16.11 \pm 0.08$ \\ 
2015-Dec-14 & 57370 &  $16.57 \pm 0.09$ & $17.24 \pm 0.08$ & $16.17 \pm 0.07$ & $15.89 \pm 0.07$ & $16.05 \pm 0.09$ & $16.15 \pm 0.07$ \\ 
2015-Dec-18 & 57374 &  $16.60 \pm 0.15$ & $16.93 \pm 0.11$ & $16.04 \pm 0.10$ & $15.84 \pm 0.10$ & $15.85 \pm 0.11$ & $16.18 \pm 0.09$ \\ 
2015-Dec-30 & 57386 &  $16.28 \pm 0.10$ & $16.67 \pm 0.08$ & $15.78 \pm 0.07$ & $15.65 \pm 0.08$ & $15.76 \pm 0.08$ & $15.98 \pm 0.07$ \\ 
2015-Dec-31 & 57387 &  $16.26 \pm 0.10$ & $16.76 \pm 0.09$ & $15.88 \pm 0.08$ & $15.62 \pm 0.08$ & $15.84 \pm 0.08$ & $15.98 \pm 0.07$ \\ 
2016-Jan-01 & 57388 &  $16.30 \pm 0.09$ & $16.66 \pm 0.07$ & $15.73 \pm 0.07$ & $15.70 \pm 0.08$ & $15.71 \pm 0.08$ & $15.97 \pm 0.07$ \\ 
2016-Jan-02 & 57389 &  $16.25 \pm 0.09$ & $16.63 \pm 0.07$ & $15.77 \pm 0.07$ & $15.65 \pm 0.08$ & $15.70 \pm 0.14$ & $15.88 \pm 0.07$ \\ 
2016-Jan-03 & 57390 &  $16.11 \pm 0.08$ & $16.63 \pm 0.08$ & $15.64 \pm 0.06$ & $15.58 \pm 0.07$ & $15.65 \pm 0.08$ & $15.85 \pm 0.07$ \\
2016-Jun-21 & 57560 &  -- & -- & $15.47 \pm 0.05$ & -- & -- & -- \\ 
2016-Jun-30 & 57569 &  -- & -- & --& -- & $15.34 \pm 0.08$ & $15.89 \pm 0.07$ \\ 
2016-Aug-24 & 57624 &  $14.60 \pm 0.04$ & $15.18 \pm 0.05$ & $14.48 \pm 0.05$ & $14.59 \pm 0.06$ & $14.78 \pm 0.06$ & $14.99 \pm 0.06$ \\ 
2016-Aug-25 & 57625 &  $14.65 \pm 0.05$ & $15.25 \pm 0.05$ & $14.47 \pm 0.05$ & $14.69 \pm 0.07$ & $14.86 \pm 0.07$ & $15.11 \pm 0.06$ \\ 
2016-Aug-26 & 57626 &  $14.87 \pm 0.05$ & $15.37 \pm 0.05$ & $14.58 \pm 0.05$ & $14.70 \pm 0.05$ & $14.91 \pm 0.07$ & $15.15 \pm 0.06$ \\
2016-Aug-27 & 57627 &  $14.98 \pm 0.05$ & $15.52 \pm 0.03$ & $14.75 \pm 0.05$ & $14.87 \pm 0.06$ & $15.10 \pm 0.07$ & $15.34 \pm 0.06$ \\ 
2016-Aug-28 & 57628 &  $14.82 \pm 0.06$ & $15.46 \pm 0.06$ & $14.67 \pm 0.06$ & $14.67 \pm 0.08$ & $15.03 \pm 0.09$ & $15.19 \pm 0.07$ \\ 
2016-Aug-29 & 57629 &  $14.61 \pm 0.05$ & $15.19 \pm 0.05$ & $14.40 \pm 0.05$ & $14.57 \pm 0.06$ & $14.75 \pm 0.07$ & $14.92 \pm 0.06$ \\ 
2016-Aug-30 & 57630 &  $15.52 \pm 0.06$ & $15.96 \pm 0.05$ & $15.23 \pm 0.06$ & $15.22 \pm 0.07$ & $15.43 \pm 0.07$ & $15.63 \pm 0.06$ \\ 
2016-Aug-31 & 57631 &  $15.46 \pm 0.05$ & $15.92 \pm 0.05$ & $15.10 \pm 0.05$ & $15.17 \pm 0.06$ & $15.35 \pm 0.07$ & $15.54 \pm 0.06$ \\ 
2016-Sep-02 & 57633 &  $15.55 \pm 0.07$ & $16.07 \pm 0.06$ & $15.29 \pm 0.06$ & $15.36 \pm 0.07$ & $15.59 \pm 0.07$ & $15.77 \pm 0.07$ \\ 
2016-Sep-02 & 57633 &  $15.31 \pm 0.06$ & $15.87 \pm 0.06$ & $15.09 \pm 0.06$ & $15.07 \pm 0.07$ & $15.22 \pm 0.21$ & $15.44 \pm 0.07$ \\ 
2016-Sep-03 & 57634 &  $15.29 \pm 0.06$ & $15.83 \pm 0.06$ & $15.07 \pm 0.06$ & $15.12 \pm 0.07$ & $15.24 \pm 0.08$ & $15.42 \pm 0.06$ \\ 
2016-Sep-04 & 57635 &  $15.32 \pm 0.06$ & $15.85 \pm 0.06$ & $15.11 \pm 0.06$ & $15.15 \pm 0.07$ & $15.27 \pm 0.08$ & $15.54 \pm 0.07$ \\ 
2016-Sep-08 & 57639 &  -- & -- & -- & $15.38 \pm 0.06$ & -- & -- \\ 
2016-Sep-12 & 57643 &  -- & -- & -- & $15.52 \pm 0.06$ & -- & -- \\ 
2016-Sep-14 & 57645 &  -- & -- & -- & -- & -- & $16.06 \pm 0.06$  \\ 
2016-Sep-17 & 57648 &  -- & -- & -- & -- & -- & $15.79 \pm 0.05$  \\ 
2016-Sep-20 & 57651 &  -- & -- & -- & $15.49 \pm 0.05$ & -- & -- \\ 
2016-Sep-26 & 57657 &  -- & -- & -- & -- & -- & $15.77 \pm 0.06$ \\ 
2016-Oct-02 & 57663 &  -- & -- & -- & $15.45 \pm 0.06$ & -- & -- \\ 
2016-Oct-08 & 57669 &  -- & -- & -- & -- & -- & $15.63 \pm 0.06$ \\ 
2016-Oct-14 & 57675 &  -- & -- & -- & $14.92 \pm 0.06$ & -- & --  \\ 
2016-Oct-20 & 57681 &  -- & -- & -- & -- & -- & $15.03 \pm 0.06$ \\ 
2016-Oct-27 & 57688 &  $14.46 \pm 0.05$ & $15.00 \pm 0.05$ & $14.26 \pm 0.05$ & $14.33 \pm 0.06$ & $14.53 \pm 0.06$ & $14.65 \pm 0.06$ \\ 
2016-Oct-28 & 57689 &  $14.37 \pm 0.05$ & $14.92 \pm 0.06$ & $14.09 \pm 0.05$ & $14.21 \pm 0.06$ & $14.38 \pm 0.07$ & $14.54 \pm 0.06$ \\ 
2016-Oct-29 & 57690 &  $14.38 \pm 0.05$ & $14.90 \pm 0.05$ & $14.13 \pm 0.05$ & $14.22 \pm 0.06$ & $14.37 \pm 0.07$ & $14.56 \pm 0.06$ \\ 
2016-Oct-30 & 57691 &  $14.76 \pm 0.05$ & $15.28 \pm 0.05$ & $14.53 \pm 0.05$ & $14.58 \pm 0.06$ & $14.79 \pm 0.06$ & $14.93 \pm 0.06$ \\ 
2016-Oct-31 & 57692 &  $14.86 \pm 0.05$ & $15.30 \pm 0.05$ & $14.54 \pm 0.05$ & $14.58 \pm 0.06$ & $14.65 \pm 0.03$ & $14.83 \pm 0.06$ \\ 
2016-Nov-14 & 57706 &  $13.98 \pm 0.04$ & $14.56 \pm 0.05$ & $13.82 \pm 0.05$ & $13.94 \pm 0.06$ & $14.15 \pm 0.06$ & $14.28 \pm 0.06$ \\
2016-Nov-16 & 57708 &  $14.07 \pm 0.05$ & $14.59 \pm 0.05$ & $13.93 \pm 0.05$ & $13.99 \pm 0.06$ & $14.17 \pm 0.07$ & $14.34 \pm 0.06$ \\ 
2016-Nov-18 & 57710 &  $14.37 \pm 0.05$ & $14.95 \pm 0.06$ & $14.15 \pm 0.06$ & $14.17 \pm 0.07$ & $14.45 \pm 0.08$ & $14.55 \pm 0.03$ \\ 
2016-Nov-20 & 57712 &  $13.81 \pm 0.05$ & $14.35 \pm 0.05$ & $13.60 \pm 0.06$ & $13.63 \pm 0.06$ & $13.85 \pm 0.07$ & $13.99 \pm 0.06$ \\ 
2016-Nov-22 & 57714 &  $13.49 \pm 0.04$ & $13.97 \pm 0.05$ & $13.13 \pm 0.05$ & $13.23 \pm 0.06$ & $13.34 \pm 0.04$ & $13.51 \pm 0.06$ \\ 
2016-Nov-23 & 57715 &  $13.32 \pm 0.04$ & $13.82 \pm 0.05$ & $13.00 \pm 0.05$ & $13.07 \pm 0.06$ & $13.21 \pm 0.03$ & $13.39 \pm 0.06$ \\ 
2016-Nov-27 & 57719 &  $13.95 \pm 0.05$ & $14.45 \pm 0.05$ & $13.70 \pm 0.05$ & $13.76 \pm 0.06$ & $13.97 \pm 0.07$ & $14.15 \pm 0.06$ \\ 
2016-Nov-30 & 57722 &  $12.87 \pm 0.05$ & $13.41 \pm 0.05$ & $12.57 \pm 0.05$ & $12.62 \pm 0.06$ & $12.77 \pm 0.07$ & $12.92 \pm 0.06$ \\ 
2016-Dec-01 & 57723 &  $12.68 \pm 0.04$ & $13.19 \pm 0.04$ & $12.40 \pm 0.05$ & $12.40 \pm 0.05$ & $12.54 \pm 0.05$ & $12.66 \pm 0.05$ \\ 
2016-Dec-06 & 57728 &  $12.89 \pm 0.04$ & $13.42 \pm 0.04$ & $12.64 \pm 0.05$ & $12.68 \pm 0.05$ & $12.80 \pm 0.06$ & $12.91 \pm 0.05$ \\ 
2016-Dec-13 & 57735 &  -- & -- & -- & $13.28 \pm 0.06$ & -- & --  \\ 
2016-Dec-16 & 57738 &  $11.74 \pm 0.04$ & $12.57 \pm 0.03$ & $11.79 \pm 0.03$ & $11.61 \pm 0.04$ & $11.82 \pm 0.06$ & $11.97 \pm 0.02$ \\ 
2016-Dec-18 & 57740 &  $13.23 \pm 0.04$ & $13.79 \pm 0.04$ & $13.07 \pm 0.05$ & $13.19 \pm 0.05$ & $13.40 \pm 0.06$ & $13.52 \pm 0.05$  \\ 
2016-Dec-20 & 57742 &  $12.26 \pm 0.04$ & $12.81 \pm 0.05$ & $12.01 \pm 0.05$ & $12.08 \pm 0.05$ & $12.30 \pm 0.06$ & $12.45 \pm 0.05$ \\ 
2016-Dec-23 & 57745 &  $12.14 \pm 0.04$ & $12.65 \pm 0.06$ & $11.83 \pm 0.05$ & $11.92 \pm 0.05$ & $12.12 \pm 0.05$ & $12.24 \pm 0.05$ \\ 
2016-Dec-26 & 57748 &  $12.72 \pm 0.04$ & $13.25 \pm 0.04$ & $12.50 \pm 0.05$ & $12.58 \pm 0.05$ & $12.84 \pm 0.06$ & $12.97 \pm 0.05$ \\ 
2016-Dec-29 & 57751 &  $11.44 \pm 0.06$ & $12.57 \pm 0.05$ & $11.80 \pm 0.05$ & $11.19 \pm 0.05$ & $11.36 \pm 0.05$ & $11.52 \pm 0.05$ \\ 
\hline
\end{tabular}
\end{center}
\end{table*}

\addtocounter{table}{-1}
\begin{table*}
\begin{center}
\begin{tabular}{cccccccc}
\hline
\multicolumn{1}{c}{Date (UT)} &
\multicolumn{1}{c}{MJD}       &
\multicolumn{1}{c}{$v$}         &
\multicolumn{1}{c}{$b$}         &
\multicolumn{1}{c}{$u$}         &
\multicolumn{1}{c}{$w1$}        &
\multicolumn{1}{c}{$m2$}        &
\multicolumn{1}{c}{$w2$}        \\
\hline
2016-Dec-30 & 57752 & $11.53 \pm 0.04$ & $12.56 \pm 0.06$ & $11.78 \pm 0.06$ & $11.33 \pm 0.04$ & $11.64 \pm 0.05$ & $11.64 \pm 0.05$ \\
2016-Dec-31 & 57753 & $11.78 \pm 0.04$ & $12.68 \pm 0.05$ & $11.91 \pm 0.05$ & $11.55 \pm 0.02$ & $11.76 \pm 0.05$ & $11.88 \pm 0.02$ \\
2017-Jan-01 & 57754 & $12.17 \pm 0.06$ & $12.75 \pm 0.05$ & $11.98 \pm 0.05$ & $12.06 \pm 0.05$ & $12.22 \pm 0.05$ & $12.37 \pm 0.05$ \\
2017-Jan-02 & 57755 & $12.92 \pm 0.04$ & $13.48 \pm 0.05$ & $12.73 \pm 0.05$ & $12.86 \pm 0.06$ & $13.07 \pm 0.06$ & $13.24 \pm 0.05$ \\
2017-Jan-06 & 57759 & $11.81 \pm 0.04$ & $12.57 \pm 0.05$ & $11.78 \pm 0.05$ & $11.54 \pm 0.05$ & $11.69 \pm 0.05$ & $11.87 \pm 0.05$ \\
2017-Jan-08 & 57761 & $12.16 \pm 0.03$ & $12.68 \pm 0.05$ & $11.86 \pm 0.05$ & $11.89 \pm 0.05$ & $12.09 \pm 0.05$ & $12.20 \pm 0.05$ \\  
2017-Jan-10 & 57763 & $12.57 \pm 0.04$ & $13.12 \pm 0.05$ & $12.30 \pm 0.05$ & $12.41 \pm 0.05$ & $12.62 \pm 0.06$ & $12.74 \pm 0.05$ \\
2017-Jan-12 & 57765 & -- & $13.78 \pm 0.05$ & $13.01 \pm 0.05$ & $13.12 \pm 0.05$ & -- & -- \\
2017-Jan-15 & 57768 & $13.33 \pm 0.04$ & $13.86 \pm 0.05$ & $13.09 \pm 0.05$ & $13.17 \pm 0.06$ & $13.22 \pm 0.06$ & $13.45 \pm 0.05$ \\
2017-Jan-18 & 57771 & $13.57 \pm 0.05$ & $14.13 \pm 0.05$ & $13.27 \pm 0.05$ & $13.36 \pm 0.06$ & $13.58 \pm 0.06$ & $13.71 \pm 0.06$ \\
\hline
\end{tabular}
\end{center}
\end{table*}

\clearpage

\setcounter{table}{2}
\begin{table*}
\caption{Log and fitting results of REM observations of CTA~102 in $J$, $H$, and $K$ bands.}
\label{REMAppendix}
\begin{center}
\begin{tabular}{ccccc}
\hline
\multicolumn{1}{c}{Date (UT)} &
\multicolumn{1}{c}{MJD} &
\multicolumn{1}{c}{$J$} &  
\multicolumn{1}{c}{$H$} &
\multicolumn{1}{c}{$K$} \\
\multicolumn{1}{c}{} &
\multicolumn{1}{c}{} &
\multicolumn{1}{c}{(mag)} &
\multicolumn{1}{c}{(mag)} &
\multicolumn{1}{c}{(mag)} \\
\hline
2016-11-15 & 57707  & --                 & $11.062 \pm 0.038$ &  $10.251 \pm 0.206$ \\
2016-11-16 & 57708  & $11.889 \pm 0.036$ & $11.089 \pm 0.042$ &  $10.209 \pm 0.071$ \\
2016-11-17 & 57709  & $11.762 \pm 0.040$ & --                 &  -- \\
2016-11-18 & 57710  & --                 & $11.181 \pm 0.038$ &  -- \\
2016-11-20 & 57712  & $11.559 \pm 0.020$ & $10.776 \pm 0.032$ &  $10.017 \pm 0.090$ \\
2016-11-21 & 57713  & $11.635 \pm 0.018$ & $10.791 \pm 0.035$ &  $10.032 \pm 0.156$ \\
2016-11-23 & 57715  & $11.167 \pm 0.036$ & $10.394 \pm 0.035$ &  -- \\
2016-11-25 & 57717  & $11.498 \pm 0.037$ & $10.671 \pm 0.059$ &  $9.938 \pm 0.151$ \\
2016-11-27 & 57719  & $11.672 \pm 0.034$ & $10.863 \pm 0.028$ &  $10.051 \pm 0.119$ \\
2016-11-29 & 57721  & $11.042 \pm 0.045$ & $10.293 \pm 0.072$ &  -- \\
2016-12-02 & 57724  & $10.510 \pm 0.034$ & $ 9.727 \pm 0.044$ &  $ 8.903 \pm 0.265$ \\
2016-12-04 & 57726  & $10.573 \pm 0.033$ & $ 9.808 \pm 0.054$ &  -- \\
2016-12-06 & 57728  & $10.811 \pm 0.057$ & $10.040 \pm 0.065$ &  -- \\
2016-12-11 & 57733  & $11.193 \pm 0.056$ & $10.345 \pm 0.048$ &  -- \\
\hline
\end{tabular}
\end{center}
\end{table*}

\clearpage

\begin{figure*}
\centering
\includegraphics[width=11cm]{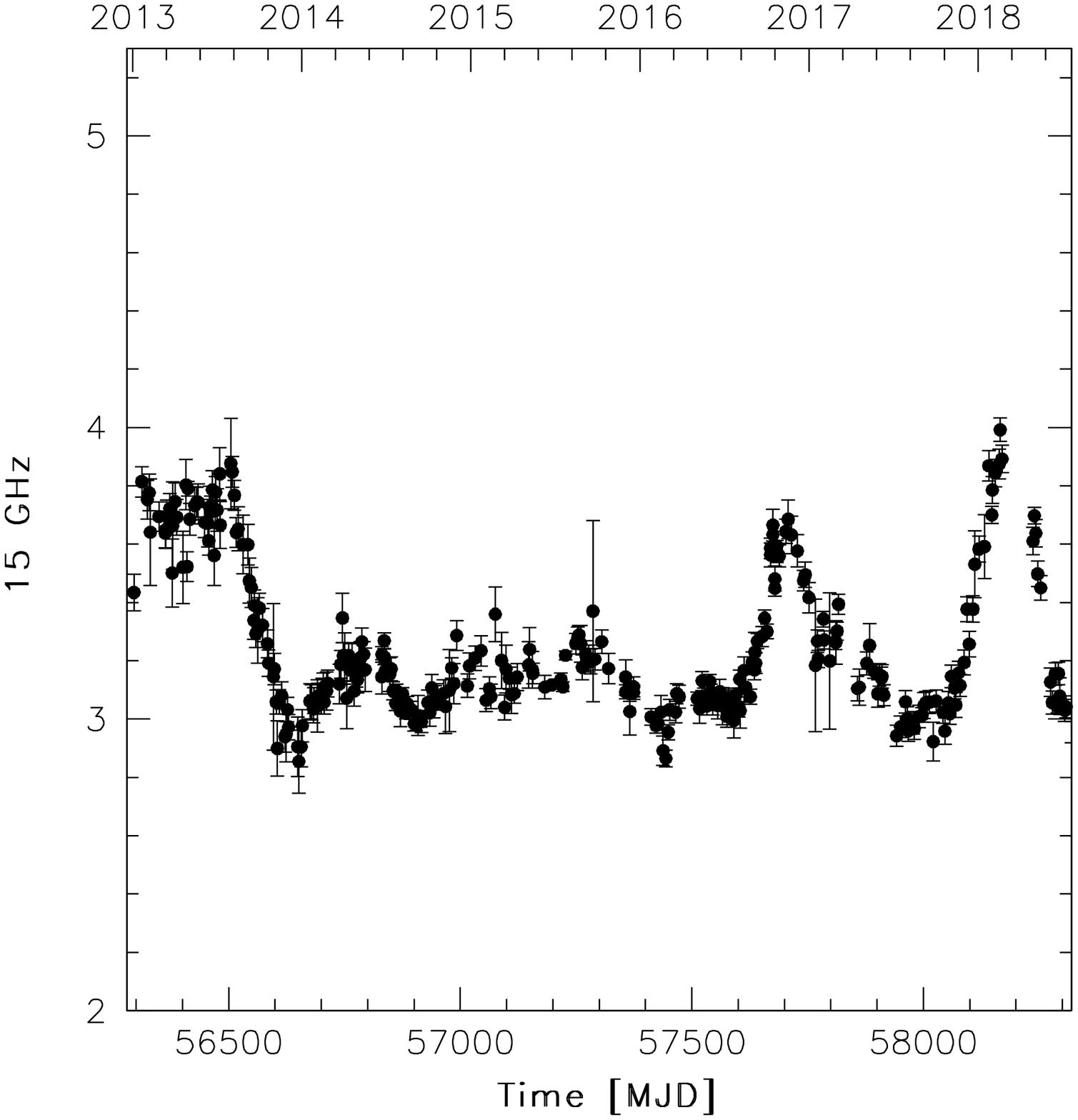}
\caption{OVRO light curve of CTA~102 at 15 GHz between 2013 January 1 (MJD 56293) and 2018 July 21 (MJD 58320).}
\label{OVRO_extended}
\end{figure*}

\bsp  
\label{lastpage}


\begin{thebibliography}{99}

\bibitem[Abdo et al.(2011)]{abdo11} Abdo, A. A., et al. 2011, ApJ, 733, L26

\bibitem[Acero et al.(2016)]{acero16} Acero, F., et al. 2016, ApJS, 223, 2

\bibitem[Ade et al.(2016)]{planck16} Ade,  P. A. R., et al. 2016, A\&A, 594, A13, 63

\bibitem[Agudo et al.(2007)]{agudo07} Agudo, I., et al. 2007, A\&A, 476, L17

\bibitem[Agudo et al.(2012)]{agudo12} Agudo, I., et al. 2012, ApJ, 747, 1

\bibitem[Aleksic et al.(2015)]{aleksic15} Aleksic, J., et al. 2015, A\&A, 578A, 22

\bibitem[Aleksic et al.(2017)]{aleksic17} Aleksic, J., et al. 2017, A\&A, 603A, 29

\bibitem[Atwood et al.(2009)]{atwood09} Atwood, W. B., et al. 2009, ApJ, 697, 1071

\bibitem[Atwood et al.(2013)]{atwood13} Atwood, W. B., et al. 2013, 2012 Fermi Symposium proceedings - eConf C121028 (arXiv:1303.3514)

\bibitem[Baars et al.(1977)]{baars77} Baars, J. W. M., Genzel, R., Pauliny-Toth, I. I. K., Witzel, A. 1977 A\&A, 61, 99

\bibitem[Barthelmy et al.(2005)]{barthelmy05} Barthelmy, S. D., et al. 2005, Space Sci. Rev., 120, 143 

\bibitem[Blandford $\&$ Rees(1978)]{blandford78} Blandford, R.~D., \& Rees, M.~J. 1978, in Pittsburgh Conference on BL Lac Objects, ed A.~M. Wolfe, University Pittsburgh Press, 328 

\bibitem[Blom et al.(1995)]{blom95} Blom, J. J., et al. 1995, A\&A, 295, 330

\bibitem[B{\"o}ttcher et al.(2013)]{boettcher13} B{\"o}ttcher, M., Reimer, A., Sweeney, K., Prakash, A. 2013, ApJ, 768, 54

\bibitem[Breeveld et al.(2010)]{breeveld10} Breeveld, A. A., et al. 2010, MNRAS, 406, 1687

\bibitem[Britzen et al.(2017)]{bri17} Britzen, S., Fendt, C., Eckart, A., \& Karas, V.\ 2017, \aap, 601, A52 

\bibitem[Britzen et al.(2018)]{bri18} Britzen, S., et al.\ 2018, \mnras, 478, 3199 

\bibitem[Bulgarelli et al.(2016)]{bulgarelli16} Bulgarelli, A., et al. 2016, The Astronomer's Telegram, 9911

\bibitem[Burrows et al.(2005)]{burrows05} Burrows, D. N., et al. 2005, Space Sci. Rev., 120, 165  

\bibitem[Casadio et al.(2015)]{casadio15} Casadio, C., et al. 2015, ApJ, 813, 1
  
\bibitem[Casadio et al.(2019)]{casadio19} Casadio, C., et al. 2019, A\&A, 622, A158 
  
\bibitem[Calcidese et al.(2016)]{calcidese16} Calcidese, P., et al. 2016, The Astronomer's Telegram, 9868

\bibitem[Carnerero et al.(2017)]{carnerero17} Carnerero et al. 2017, MNRAS, 450, 2677

\bibitem[Chatterjee et al.(2013)]{chatterjee13} Chatterjee, R., et al. 2013, ApJ, 763L, 11

\bibitem[Ciprini et al.(2016)]{ciprini16} Ciprini, S. 2016, The Astronomer's Telegram, 9869

\bibitem[Covino et al.(2004)]{covino04} Covino, S., Stefanon, M., Sciuto, G., et al. 2004, SPIE, 5492, 1613

\bibitem[D'Ammando et al.(2011)]{dammando11} D'Ammando, F., et al. 2011, A\&A, 529A, 145

\bibitem[D'Ammando et al.(2013)]{dammando13} D'Ammando, F., et al. 2013, MNRAS, 431, 2481

\bibitem[Edelson \& Krolik(1988)]{edelson88} Edelson, R. A., \& {Krolik}, J. H.  1988, ApJ, 333, 646

\bibitem[Fossati et al.(2008)]{fossati08} Fossati, G., et al. 2008, ApJ, 677, 906

\bibitem[Fromm et al.(2013)]{fro13} Fromm, C.~M., et al.\ 2013, \aap, 557, A105 

\bibitem[Fuhrmann et al.(2014)]{fuhrmann14} Fuhrmann, L., et al. 2014, MNRAS, 441, 1899

\bibitem[Gasparyan et al.(2018)]{gasparyan18} Gasparyan S., Sahakyan, N., Baghmanyan, V., Zargaryan, D. 2018, ApJ, 863, 114

\bibitem[Gehrels et al.(2004)]{gehrels04} Gehrels, N., et al. 2004, ApJ, 611, 1005 

\bibitem[Ghisellini et al.(2011)]{ghisellini11} Ghisellini, G., Tavecchio, F.,  Foschini, L., Ghirlanda, G. 2011, MNRAS, 414, 2674

\bibitem[Godfrey \& Shabala(2013)]{godfrey13} Godfrey,  L. E. H., Shabala, S. S. 2013, 767, 12, 9
  
\bibitem[Hartman et al.(1999)]{hartman99} Hartman, R. C., et al. 1999, ApJS, 123, 79

\bibitem[Hayashida et al.(2012)]{hayashida12} Hayashida, M., et al. 2012, ApJ, 754, 114
  
\bibitem[Hayashida et al.(2015)]{hayashida15} Hayashida, M., et al. 2015, ApJ, 807, 79 
 
\bibitem[Hufnagel \& Bregman(1992)]{hufnagel92} Hufnagel, B. R., \& Bregman, J. N.  1992, ApJ, 386, 473

\bibitem[Kalberla et al.(2005)]{kalberla05} Kalberla, P. M. W., et al. 2005, A$\&$A, 440, 775

\bibitem[Katajainen et al.(2000)]{katajainen00} Katajainen S., et al., 2000, A\&AS, 143, 357

\bibitem[Krawczynski et al.(2004)]{krawczynski04} Krawczynski, H., et al. 2004, ApJ, 601, 151
  
\bibitem[Larionov et al.(2016)]{lar16} Larionov, V.~M., et al.\ 2016, \mnras, 461, 3047 

\bibitem[Larionov et al.(2017)]{lar17} Larionov, V.~M., et al.\ 2017, Galaxies, 5, 91

\bibitem[Liska et al.(2018)]{lis18} Liska, M., et al.\ 2018, \mnras, 474, L81 

\bibitem[Marscher et al.(2014)]{marscher14} Marscher A. 2014, ApJ, 780, 87
   
\bibitem[Mattox et al.(1996)]{mattox96} Mattox, J. R., et al. 1996, ApJ, 461, 396 

\bibitem[Merloni et al.(2007)]{merloni07} Merloni A., Heinz, S. 2007, MNRAS, 381, 589
 
\bibitem[Meyer et al.(2019)]{meyer19} Meyer, M., Scargle, J. D., Blandford, R. D. 2019, ApJ, submitted [arXiv:1902.02291]

\bibitem[Mignone et al.(2010)]{mig10} Mignone, A., Rossi, P., Bodo, G., Ferrari, A., \& Massaglia, S.\ 2010, \mnras, 402, 7 

\bibitem[Miller-Jones et al.(2019)]{mil19} Miller-Jones, J.~C.~A., et al.\ 2019, Nature, published online on 29 April 2019

\bibitem[Mingaliev et al.(2001)]{mingaliev01} Mingaliev, M. G., et al. 2001, A\&A, 370, 78

\bibitem[Moretti et al.(2005)]{moretti05} Moretti, A., et  al.,  2005,  in  Siegmund  O.  H.  W.,  ed.,  Proc.  SPIE  Conf.
Ser. Vol. 5898, UV, X-Ray, and Gamma-Ray Space Instrumentation for Astronomy XIV. SPIE, Bellingham, 360

\bibitem[Nakamura et al.(2001)]{nak01} Nakamura, M., Uchida, Y., \& Hirose, S.\ 2001, \na, 6, 61 

\bibitem[Narayan \& Piran (2012)]{narayan12} Narayan R., Piran T., MNRAS, 2012, 420, 604
   
\bibitem[Oh et al.(2018)]{oh18} Oh, K. et al. 2018, ApJS, 235, 4

\bibitem[Ojha et al.(2016)]{ojha16} Ojha, R., Carpenter, B., D'Ammando, F. 2016, The Astronomer's Telegram, 9924

\bibitem[Orienti et al.(2019)]{orienti19} Orienti M., D’Ammando F., Giroletti M., Dallacasa D., Giovannini G., Ciprini S., 2019, MNRAS, in press (arXiv:1910.08568)

\bibitem[Osterman Meyer(2009)]{osterman09} Osterman Meyer A., et al. 2009, AJ ,138, 1902

\bibitem[Perucho et al.(2012)]{per12} Perucho, M., Mart{\'{\i}}-Vidal, I., Lobanov, A.~P., \& Hardee, P.~E.\ 2012, \aap, 545, A65

\bibitem[Peterson et al.(1998)]{peterson98} Peterson B. M.,  Wanders I., Horne K., Collier S., Alexander T., Kaspi S., Maoz D. 1998, PASP, 110, 660

\bibitem[Petropoulou et al.(2016)]{petropoulou16} Petropoulou, M., Giannios, D., Sironi, L. 2016, MNRAS, 462, 3
  
\bibitem[Pica et al.(1988)]{pica88} Pica, A. J., et al. 1988, AJ, 96, 1 

\bibitem[Pushkarev et al.(2010)]{pushkarev10} Pushkarev A. B., Kovalev Y. Y., Lister M. L., ApJ, 2010, 722 L7
  
\bibitem[Poole et al.(2008)]{poole08} Poole, T. S., et al. 2008, MNRAS, 383, 627

\bibitem[Raiteri et al.(2003)]{raiteri03} Raiteri, C.~M., et al. 2003, A\&A, 402, 151
 
\bibitem[Raiteri et al.(2010)]{raiteri10} Raiteri, C.~M., et al. 2010, A\&A, 524, 43

\bibitem[Raiteri et al.(2011)]{raiteri11} Raiteri, C.~M., et al. 2011, A\&A, 534A, 87

\bibitem[Raiteri et al.(2012)]{raiteri12} Raiteri, C.~M., et al. 2012, A\&A, 545A, 48

\bibitem[Raiteri et al.(2013)]{raiteri13} Raiteri, C.~M., et al. 2013, MNRAS, 436, 1530

\bibitem[Raiteri et al.(2014)]{raiteri14} Raiteri, C.~M., et al. 2014, MNRAS, 442, 629

\bibitem[Raiteri et al.(2017)]{rai17} Raiteri, C.~M., et al.\ 2017, \nat, 552, 374 

\bibitem[Readhead et al.(1989)]{readhead89} Readhead, A. C. S., et al. 1989, ApJ, 346, 566

\bibitem[Richards et al.(2011)]{richards11} Richards, J. L., et al. 2011, ApJS, 194, 29 

\bibitem[Righini et al.(2016)]{righini16} Righini, S., et al. 2016, The Astronomer's Telegram, 9884

\bibitem[Roming et al.(2005)]{roming05} Roming, P. W. A., et al. 2005, Space Sci. Rev., 120, 95 

\bibitem[Schmidt(1965)]{schmidt65} Schmidt, M. 1965, ApJ, 141, 1295

\bibitem[Shukla et al.(2018)]{shukla18} Shukla, A., et al. 2018, ApJ, 854, L26

\bibitem[Stickel et al.(1991)]{stickel91} Stickel M., et al. 1991, ApJ, 374, 431

\bibitem[Villata et al.(1997)]{villata97} Villata, M., et al. 1997, A\&AS, 121, 119

\bibitem[Wehrle et al.(1998)]{wehrle98} Wehrle, A. E., et al. 1998, ApJ, 497, 178

\bibitem[Wilms et al.(2000)]{wilms00} Wilms, J., Allen, A., McCray, R. 2000, ApJ, 542, 914 

\bibitem[Xu et al.(2016)]{xu16} Xu, Z.-L., et al. 2016, The Astronomer's Telegram, 9901

\bibitem[Zacharias et al.(2017)]{zacharias17} Zacharias, M., et al. 2017, ApJ, 851, 72

\bibitem[Zacharias et al.(2019)]{zacharias19} Zacharias, M., et al. 2019, ApJ, 871, 19

\bibitem[Zdziarski \& Boettcher(2015)]{zdziarski15} Zdziarski, A. A., Boettcher, M. 2015, MNRAS, 450, L21
  
\bibitem[Zerbi et al.(2001)]{zerbi01} Zerbi, R. M., Chincarini, G., Ghisellini, G., et al. 2001, AN, 322, 275

\end{thebibliography}
\end{document}